\documentclass[12pt,a4paper]{article}

\usepackage{latexsym}
\usepackage{amssymb,amsmath}
\usepackage{graphicx}
\usepackage{ae}
\usepackage[utf8]{inputenc}
\usepackage[T1]{fontenc}
\usepackage{amsmath,amsthm,rotating}
\usepackage{colortbl}
\usepackage{tikz,xtab}
\usepackage{moreverb}
\usepackage{alltt}
\usepackage{helvet}
\usepackage{amsmath}
\usepackage{algorithm2e, setspace}
\usetikzlibrary{decorations.pathreplacing}

\newcommand\smallmath[2]{#1{\raisebox{\dimexpr \fontdimen 22 \textfont 2
      - \fontdimen 22 \scriptfont 2 \relax}{$\scriptstyle #2$}}}

\def\f{\frac}
\def\eot{\hfill{$\Box$}}
\newcommand{\lettnum}{\def\labelenumi{(\alph{enumi})}}

\newcommand{\la}{\(\leftarrow\ \)}

\newcommand\smalloplus{\smallmath\mathbin\oplus}

\newcommand{\comm}[1]{}

\newcommand{\smallsf}[1]{\small{\textsf {#1}}}
\newcommand{\bhash}{\ensuremath{\mathbf b^{\#}}}
\newtheorem{thm}{Theorem}[section]
\newtheorem{lemma}[thm]{Lemma}

\newtheorem{cor}[thm]{Corollary}

\begin{document}

\vspace{-10mm}
\title{Fast Generation of Unlabelled Free Trees\\using Weight Sequences}
\author{Paul Brown\footnote{paulb@dcs.bbk.ac.uk} $\,$ and 
Trevor Fenner\footnote {trevor@dcs.bbk.ac.uk}
\\Department of Computer Science and Information Systems
\\Birkbeck, University of London, London WC1E 7HX, U.K.}
\maketitle
\vspace{-10mm}
\begin{abstract}
{\rm In this paper, we introduce a new representation for {\em ordered} trees, the {\em weight sequence} representation. We then use this to construct new representations for both {\em rooted} trees and {\em free} trees, namely the {\em canonical weight sequence} representation. We construct algorithms for generating the weight sequence representations for all rooted and free trees of order $n$, and then add a number of modifications to improve the efficiency of the algorithms. Python implementations of the algorithms incorporate further improvements by using generators to avoid having to store the long lists of trees returned by the recursive calls, as well as caching the lists for rooted trees of small order, thereby eliminating many of the recursive calls.  We further show how the algorithm can be modified to generate adjacency list and adjacency matrix representations for free trees. We compared the runtimes of our Python implementation for generating free trees with the Python implementation of the well-known WROM algorithm taken from NetworkX. The implementation of our algorithm is over four times as fast as the implementation of the WROM algorithm. The runtimes for generating adjacency lists and matrices are somewhat longer than those for weight sequences, but are still over three times as fast as the corresponding implementations of the WROM algorithm.}
\end{abstract}
\vspace{-2.5mm}
\section{\normalsize Introduction}\label{introduction}\vspace{-2.5mm}
The enumeration of {\em trees}, whether {\em ordered}, {\em rooted} or {\em free}, has been well-studied. (Rooted trees are also called {\em oriented} trees, see \cite{knut}). Indeed, ``Cayley's formula'', which states that there are precisely $n^{n-2}$ free trees on $n$ {\em labelled} vertices, dates back to Carl Wilhelm Borchard in the middle of the nineteenth century \cite{borc}. In $1948$, Otter \cite{otte} derived asymptotic estimates for the numbers of both {\em unlabelled} free and rooted trees. In addition, generating functions for the numbers of both unlabelled free and rooted trees have been obtained (see \cite{knut}). The exact counts for unlabelled free trees with $n$ vertices, for $n \le 36$, are listed as Sequence $A000055$ in the OEIS \cite{oeis}.

One of the first efficient algorithms for {\em generating} unlabelled rooted trees was developed by Beyer and Hedetniemi \cite{baye} using a {\em level sequence representation}. This algorithm was extended by Wright, Richmond, Odlyzko and McKay \cite{wrom} to generate all unlabelled free trees. This algorithm is referred to informally as the WROM algorithm, and is by far the most commonly used algorithm to generate non-isomorphic free trees. An alternative algorithm was constructed by Li and Ruskey \cite{lirusk} using the {\em parent sequence representation}. Indeed, a good survey of this topic can be found in Li's thesis \cite{li}; see also \cite{knut2}. Other work in this area has recently been conducted by Sawada \cite{sawa}, who presented algorithms to generate both rooted and free {\em plane} (i.e.,  ordered) trees. 


In this paper, we construct new recursive algorithms for generating rooted trees and free trees, taking a different approach from previous authors. We introduce a new representation, the {\em canonical weight sequence}, and use this rather than level or parent sequences. Unlike the latter representations, the weight sequence representation preserves {\em referential transparency} for subtrees. This makes possible major improvements in efficiency by enabling us to efficiently cache the weight sequences of rooted trees of small order, thereby eliminating many of the recursive calls. 

We introduce a number of modifications to improve the efficiency of the basic algorithms. We implemented the algorithms in Python, incorporating further improvements by using generators to avoid having to store the long lists of trees returned by the recursive calls. We also show how the algorithm can be amended to generate adjacency list and adjacency matrix representations for free trees.

We compared the runtimes of our Python implementation for generating free trees with the Python implementation of the well-known WROM algorithm taken from NetworkX \cite{nx}, the popular Python graph and network libary. The Python implementation of our new algorithm is over four times as fast as the corresponding implementation of the WROM algorithm. 

The programs were all written in Python $3.7$ and executed using the PyPy$3$ compiler, although the pseudo-code we present can easily be translated into other languages. The Python code can be found in the appendix.  Any graph-theoretic terminology and notation not explicitly defined can be found in Bondy and Murty's text \cite{b&m1}.

In Section \ref{preliminaries}, we introduce the weight sequence representations for ordered trees, weighted trees and free trees. In Sections \ref{rooted tree generation} and \ref{free tree generation}, we present our algorithms for generating rooted and free trees, respectively. Then, in Section \ref{code efficiencies}, we discuss improvements to the algorithms and their implementations, as well as the modifications required to generate the adjacency list and matrix representations of the trees. In Section \ref{time tests}, we compare the runtimes of the Python implementations of our algorithm with those of the WROM algorithm, and Section \ref{conclusion} contains our concluding remarks.
\section{\normalsize Preliminaries}\label{preliminaries}\vspace{-2.5mm}
\subsection{\normalsize Notation}\label{notation}\vspace{-2.5mm}
A {\em free tree} $T$ is an connected undirected graph that contains no cycles (conventionally, just called a {\em tree} in the graph theory literature). The {\em degree} of a vertex $v$ of $T$ is the number of vertices adjacent to $v$. A {\em leaf} of $T$ is a vertex of degree $1$; all other vertices of $T$ are called {\em branch} vertices. It is easy to show that there is a unique path between any pair of vertices of $T$.
\newpage
A {\em rooted tree} $R$ is a free tree with a distinguished vertex called its {\em root}. Let $v$ be a vertex of $R$. Any other vertex $u$ on the path from the root to $v$ is an {\em ancestor} of $v$, and $v$ is a {\em descendant} of $u$. A descendant $w$ of $v$ that is adjacent to $v$ is a {\em child} of $v$, and $v$ is the {\em parent} of $w$. Any other child of $v$ is a {\em sibling} of $w$. By definition, the root has no parent.

Let $v$ be any descendant of the root of $R$. The {\em subtree} of $R$ that consists of $v$ together with all of its descendants can clearly be considered to be a rooted tree with root $v$. We denote this subtree by $R(v)$ and define $wt(v)$, the {\em weight} of $v$, to be the order of $R(v)$; so the weight of the root of $R$ is the order of $R$. If $v$ is a leaf then $wt(v)=1$ and $R(v)$ contains just the vertex $v$. $R - R(v)$ is the rooted tree, with the same root as $R$, obtained from $R$ by deleting the subtree $R(v)$ together with the edge between $v$ and its parent.

An {\em ordered tree}, sometimes called a {\em plane tree} \cite{sawa}, is a rooted tree in which there is an ordering defined on the children of each vertex. By convention, when drawing a rooted tree, the root is placed at the top of the diagram and, for an ordered tree, the order of the children is from left to right. So we may refer to the {\em first} (left-most) or {\em last} (right-most) child of its parent. Similarly, for any vertex $v$ that is not the last child of its parent, we may refer to the {\em next} sibling of $v$. We note that, if $R$ is an ordered tree, the subtrees $R(v)$ and $R-R(v)$ are considered to be ordered trees, inheriting the ordering of the sets of children from $R$. For convenience, when $w$ is a child of $v$, instead of saying that $R(w)$ is subtree of $R(v)$, we often say that $R(w)$ is a subtree of $v$.

A tree is called a {\em labelled tree} if each vertex is assigned a unique label. For any unlabelled ordered tree $R$ with $n$ vertices, we conventionally label the vertices as $v_1,\, v_2, \ldots ,v_n$ in {\em pre-order}, where $v_1$ is the root of the tree. {\em Pre-order} is the total ordering of the vertices of $R$ defined recursively as follows: for any vertex $u$ with children $u_1, \, u_2, \ldots, u_p$, {\em pre-order} for the subtree $R(u)$ starts with $u$, followed by the vertices of $R(u_1)$ in pre-order (if $p \ge 1$), then the vertices of $R(u_2)$ in pre-order (if $p \ge 2$), etc. We note that $v_2$ is the first child of the root $v_1$. It trivially follows that, for any vertex $v_k$ of $R$, the pre-order of the vertices of $R(v_k)$ is a contiguous subsequence of the pre-order of the vertices of $R$. 

Two labelled free trees are {\em isomorphic} if there is a bijection between their vertex sets that preserves adjacency and non-adjacency; two labelled rooted trees are isomorphic if there exists an isomorphism between their underlying free trees that {\em maps the root of one onto root of the other}; two labelled ordered trees are isomorphic if there exists an isomorphism between their underlying rooted trees {\em that preserves the orderings of the children of each vertex}. We say that two trees (whether ordered, rooted or free) are {\em f-isomorphic} if their underlying free trees are isomorphic, and that two trees (whether ordered or rooted) are {\em r-isomorphic} if their underlying rooted trees are isomorphic. For completeness, we will also say that two isomorphic ordered trees are {\em o-isomorphic}.

An {\em integer sequence} $\bf s$ is an (ordered) list of integers $s_1\, s_2 \dots s_n$. In this paper, we shall assume that every element $s_i$ in $\bf s$ is positive, and denote the length of $\bf s$ by $|\bf s|$; so in this case $|{\bf s}|=n$. If ${\bf t} = t_1\, t_2 \dots t_m$ is another integer sequence, we denote the {\em concatenation} of the two sequences by ${\bf s} \smalloplus {\bf t}$, i.e., ${\bf s} \smalloplus {\bf t} =  s_1\, s_2 \dots s_n \,  t_1\, t_2 \dots t_m$. For simplicity, we do not distinguish between a sequence of length one and single integer, e.g., we may write $s_1 \smalloplus {\bf t}$.
\newpage
We say that $\bf s$ is {\em lexicographically greater than or equal to} $\bf t$, denoted $\bf s \ge \bf t$, if and only if either (a) or (b) below hold:
\begin{enumerate}\lettnum\vspace{-2.5mm}
\item $s_i = t_i$, for $1 \le i < j$, and $s_j > t_j$, for some $j$, $1 \le j \le \min(n,\, m)$;\vspace{-2.5mm}
\item $s_i = t_i$ for $1 \le i \le \min(n,\, m)$ and $n \ge m$.\vspace{-2.5mm}
\end{enumerate}
Strict lexicographical inequality ${\bf s} > {\bf t}$ holds if ${\bf s} \ge {\bf t}$ and ${\bf s} \ne {\bf t}$. We note that this defines a total ordering on the set of integer sequences.
\vspace{-2.5mm}
\subsection{\normalsize Weight sequences of ordered trees}\label{weight rep}\vspace{-2.5mm}
A common way to represent an ordered tree is by a suitable integer sequence obtained by traversing the tree in some specified order (usually pre-order) and recording some particular property of each vertex as it is visited. The resulting sequence is called a {\em representation sequence} for the tree. A {\em valid} representation for ordered trees is a representation by integer sequences such that any two ordered trees that have the same representation sequence are $o$-isomorphic.
\begin{figure}[h]
\begin{centering}
\begin{tikzpicture}[sibling distance=5em, level distance = 2.5em, 
  every node/.style = {shape=circle, draw,  inner sep=0pt, text width=6mm, align=center}]
  \node {$v_1$}
    child { node {$v_2$}	    [sibling distance=3.5em]
      child { node {$v_3$}    [sibling distance=2.5em]
        child { node {$v_4$} }
        child { node {$v_5$} }
        child { node {$v_6$} } }
      child { node {$v_7$} } }
 child { node {$v_8$}
    child { node {$v_9$} } }
    child { node {$v_{10}$} };
\end{tikzpicture}

\vspace{2.5mm}
\caption{An ordered tree rooted at $v_1$ labelled in preorder.}\label{fig 1}
\end{centering}
\end{figure}
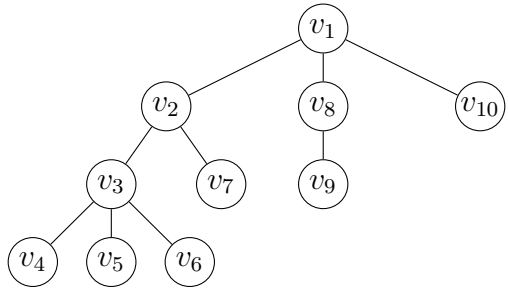

For example, consider the ordered tree of order $10$ shown in Figure $1$, in which the vertices are labelled in pre-order. If we record the {\em level} (where we define the level of the root to be $1$, the level of its children to be $2$, etc.) of each vertex in a pre-order traversal, we obtain the following sequence: $1\,2\,3\,4\,4\,4\,3\,2\,3\,2$. This is called the {\em level sequence} of the tree. Similarly, if we record the index of the label of the parent of each vertex, we obtain its {\em parent sequence}: $1\,2\,3\,3\,3\,2\,1\,8\,1$ (note there is no parent for the root in the parent sequence representation). Both of these sequence representations are well-known and have been shown to be valid representations for ordered trees (see \cite{baye} \cite{cook}). They have been used in the design of algorithms for generating rooted trees and free trees by Beyer and Hedetniemi \cite{baye}, Wright et al. \cite{wrom}, Li and Ruskey \cite{lirusk},  Sawada \cite{sawa} and Cook \cite{cook}.

In this paper, we introduce a new representation sequence. This is constructed by recording the {\em weight} of each vertex in a pre-order traversal of the tree. We call this representation the {\em weight sequence} of the tree, and denote the weight sequence of any ordered tree $R$ by ${\bf ws}(R)$. For example, for the tree $R$ in Figure $1$, ${\bf ws}(R) = 10\,\,6\,4\,1\,1\,1\,1\,2\,1\,1$.
\newpage
\begin{lemma}\label{R_k  lemma 1}
{\rm Let $R$ be an ordered tree of order $n$ with weight sequence ${\bf ws}(R) = s_1 \,s_2 \ldots s_n$, where the vertices are labelled $v_1,\, v_2, \ldots ,v_n$ in pre-order. Then
\begin{enumerate}\lettnum\vspace{-4mm}
\item $s_1 = n$;\vspace{-2.5mm}
\item for all vertices $v_k$ of $R$,
\begin{enumerate}\vspace{-2.5mm}
\item[(i)] $s_k = wt(v_k)$, i.e., the order of $R(v_k)$;\vspace{-1mm}
\item[(ii)] ${\bf ws}(R) = {\bf x} \smalloplus{\bf ws}(R(v_k)) \smalloplus {\bf y}$ for some integer sequences ${\bf x}$ and ${\bf y}$;\vspace{-1mm}
\item[(iii)] ${\bf ws}(R(v_k)) = s_k \,s_{k+1} \ldots s_{k + s_{k} -1}$;\vspace{-1mm}
\end{enumerate}
\item ${\bf ws}(R - R(v_2)) = t\,  s_{{s_{2} + 2}} \,  s_{{s_{2}+3}} \ldots s_n$ where $t = n-s_2$.\vspace{-2.5mm}
\end{enumerate}
\textit{Proof}
(a), (b)(i) and (b)(ii) follow immediately since the vertices of $R$ are labelled in pre-order.

(b)(iii) The weight of any vertex in $R(v_k)$ is the same as in $R$. Since $R(v_k)$ is of order $s_k$, it then follows that ${\bf ws}(R(v_k)) = s_k \,s_{k+1} \ldots s_{k + s_{k} -1}$.

(c) Since $R(v_2)$ is of order $s_2$, the result follows easily from (b)(iii).
\hfill\eot}
\end{lemma}
\begin{cor}\label{R_k  cor 2}
{\rm Let $R$ be an ordered tree of order $n$. Suppose that $u_1, \, u_2, \ldots, u_p$ are the children of the root of $R$, where $u_{i+1}$ is the next sibling of $u_i$ for all $i$, $1 \le i \le p-1$. Then
\begin{equation}
{\bf ws}(R) = n \smalloplus {\bf ws}(R(u_1)) \smalloplus {\bf ws}(R(u_2)) \smalloplus \ldots \smalloplus {\bf ws}(R(u_p)).\label{R_k  cor 2 eqn}
\end{equation}
}
\end{cor}
It therefore follows from Lemma \ref{R_k  lemma 1}(b) that, for any ordered tree $R$, the subsequence of ${\bf ws}(R)$ that corresponds to $R(v_k)$ is just ${\bf ws}(R(v_k))$, where $R(v_k)$ is considered as an ordered tree in its own right. This is the main reason why the weight sequence is a particularly useful representation for the generation of trees of order $n$: we can construct the weight sequence of any ordered tree of order $n$ directly from the weight sequences of its subtrees. So, if $r$ is the order of $R(u_1)$, it follows from Lemma \ref{R_k  lemma 1} and Corollary \ref{R_k  cor 2} that one way to accomplish this is to take the weight sequence of an ordered tree of order $r$ (corresponding to ${\bf ws}(R(u_1))$), and combine it appropriately with the weight sequence of an ordered tree of order $n-r$ (corresponding to ${\bf ws}(R-R(u_1))$). We shall elaborate on this in Sections \ref{rooted tree generation} and \ref{free tree generation}.

We note that, since the weight sequence of a tree is well defined, any $o$-isomorphic trees must have the same weight sequence.
\begin{lemma}\label{ord rep lemma}
{\rm The weight sequence is a valid representation for ordered trees. 

\textit{Proof} By inspection, the result clearly holds when the order is less than four. So suppose that the result holds for all ordered trees of order less than $n$, where $n \ge 4$. Let $R$ and $R^\prime$ be labelled ordered trees of order $n$ such that ${\bf ws}(R)={\bf ws}(R^\prime)$, where the vertices of the trees are labelled $v_1,\, v_2, \ldots ,v_n$ and $v^\prime_1,\, v^\prime_2, \ldots ,v^\prime_n$, respectively, in pre-order. 

Consider the subtrees $R(v_2)$ and $R - R(v_2)$ of $R$, and $R^\prime(v^\prime_2)$ and $R^\prime - R^\prime(v^\prime_2)$ of $R^\prime$. By parts (b)(iii) and (c) of Lemma \ref{R_k  lemma 1}, ${\bf ws}(R(v_2))={\bf ws}(R^\prime(v^\prime_2))$ and ${\bf ws}(R - R(v_2))={\bf ws}(R^\prime - R^\prime(v^\prime_2))$. Since these trees are of order less than $n$, it follows from the inductive hypothesis that $R(v_2)$ is $o$-isomorphic to $R^\prime(v^\prime_2)$, and $R - R(v_2)$ is $o$-isomorphic to $R^\prime - R(v^\prime_2)$. So, since $v_2$ and $v^\prime_2$ are the first children of the roots of $R$ and $R^\prime$, respectively, it follows that $R$ is $o$-isomorphic to $R^\prime$. Hence the weight sequence is a valid representation for ordered trees.
\hfill\eot}
\end{lemma}
The following lemma will be used in Section \ref{can form}.
\begin{lemma}\label{int seq lemma}
{\rm Let $\bf s$ and $\bf t$ be weight sequences of trees. If ${\bf s} > {\bf t}$ then ${\bf x} \smalloplus {\bf s} \smalloplus {\bf y} > {\bf x} \smalloplus {\bf t} \smalloplus {\bf z}$, for any integer sequences ${\bf x}$, ${\bf y}$ and ${\bf z}$.

\textit{Proof} This follows immediately from Lemma \ref{R_k  lemma 1}(a) and the definition of lexicographical order.

\vspace{-2.5mm}
\hfill\eot}
\end{lemma}
\vspace{-2.5mm}
\subsection{\normalsize Canonical  weight sequences of rooted trees}\label{can form}\vspace{-2.5mm}
We extend the definition of a valid representation by integer sequences to rooted trees: a {\em valid} representation for rooted trees is a well-defined representation such that any two rooted trees that have the same representation sequence are $r$-isomorphic.

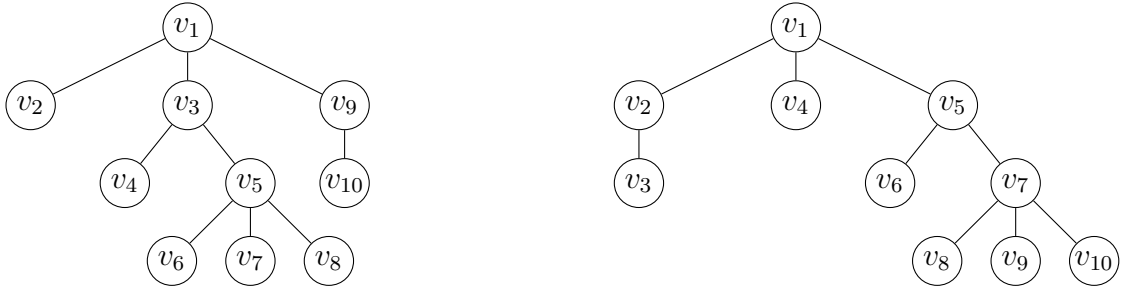
\begin{figure}[h]
\begin{centering}
\begin{tikzpicture}[sibling distance=5em, level distance = 2.5em, 
  every node/.style = {shape=circle, draw,  inner sep=0pt, text width=6mm, align=center}]
\begin{scope}[shift={(-4,0)}]
  \node {$v_1$}
    child { node {$v_2$} }
    child { node {$v_3$}     [sibling distance=4em]
      child { node {$v_4$} } 
      child { node {$v_5$}     [sibling distance=2.5em]
        child { node {$v_6$} }
        child { node {$v_7$} }
        child { node {$v_8$} } }}
    child { node {$v_9$}
    child {node {$v_{10}$} } };
\end{scope}
\begin{scope}[shift={(4, 0)}]
  \node {$v_1$}
    child { node {$v_2$}
    child { node {$v_3$} } }
    child { node {$v_4$} }
    child { node {$v_5$}     [sibling distance=4em]
      child { node {$v_6$} } 
      child { node {$v_7$}     [sibling distance=2.5em]
        child { node {$v_8$} }
        child { node {$v_9$} }
        child { node{$v_{10}$}} }};
\end{scope}
\end{tikzpicture}

\vspace{2.5mm}
\caption{Two trees that are $r$-isomorphic but not $o$-isomorphic.}\label{fig 2}
\end{centering}
\end{figure}

Now, since the weight sequence is a valid representation for ordered trees by Lemma \ref{ord rep lemma}, two $r$-isomorphic ordered trees that are not $o$-isomorphic must have different weight sequences. For example, the two $r$-isomorphic ordered trees in Figure \ref{fig 2} have weight sequences $10\,\, 1\,6\,1\, 4\,1\,1\,1\,2\,1$ and $10\,\, 2\,1\,1\,6\, 1\, 4\,1\,1\,1$, respectively (they are also $r$-isomorphic but not $o$-isomorphic to the ordered tree in Figure \ref{fig 1}). So, in order to define a valid representation for rooted trees using weight sequences, we need to choose a unique representative from each  $r$-isomorphism class of ordered trees.

An ordered tree $R$ of order $n$ is {\em canonically ordered} if ${\bf ws}(R(u)) \ge {\bf ws}(R(v))$, for each vertex $u$ of $R$ having a next sibling $v$. Clearly, if $R$ is canonically ordered then so is $R(v)$, for each vertex $v$ of $R$. It is easy to see that the ordered tree in Figure \ref{fig 1} is canonically ordered, but those in Figure \ref{fig 2} are not.

\begin{lemma}\label{cws well defined lemma}
{\rm Let $R$ and $R^\prime$ be $r$-isomorphic canonically ordered trees. Then ${\bf ws}(R) = {\bf ws}(R^\prime)$, and $R$ and $R^\prime$ are $o$-isomorphic.

\textit{Proof} Let $n$ be the order of $R$ and $R^\prime$. It is easy to see, by inspection, that the result holds when $n \le 3$. So suppose that $n \ge 4$ and that the result holds for all pairs of trees of order less than $n$. Let $u_1, \, u_2, \ldots, u_p$ be the children of the root of $R$, where $u_{i+1}$ is the next sibling of $u_i$ for each $u_i$. Let $\theta$ be an $r$-isomorphism from $R$ to $R^\prime$. Clearly, $\theta$ maps the children of the root of $R$ to the children of the root of $R^\prime$. So the subtrees of the root of $R^\prime$ are precisely the subtrees $\theta(R(u_i))$ in some order. 

For each $u_i$, both $R(u_i)$ and $\theta(R(u_i))$ are canonically ordered. Therefore, since $R(u_i)$ is $r$-isomorphic to $\theta(R(u_i))$, it follows that $R(u_i)$ is $o$-isomorphic to $\theta(R(u_i))$ by the inductive hypothesis. Hence ${\bf ws}(R(u_i)) = {\bf ws}(\theta(R(u_i)))$ for each $u_i$. Since $R$ and $R^\prime$ are both canonically ordered, it is now easy to see from (\ref{R_k  cor 2 eqn}) that ${\bf ws}(R) = {\bf ws}(R^\prime)$. Hence $R$ and $R^\prime$ are $o$-isomorphic by Lemma \ref{ord rep lemma}. 
\hfill\eot}
\end{lemma}
Clearly, for any ordered tree $R$, by suitably permuting the subtrees of each vertex, we can obtain a canonically ordered tree that is $r$-isomorphic to $R$. We therefore define ${\bf cws}(R)$, the {\em canonical weight sequence} of $R$, to be the weight sequence of any canonically ordered tree that is $r$-isomorphic to $R$. By Lemma \ref{cws well defined lemma}, ${\bf cws}(R)$ is well defined. 
\newpage
\begin{lemma}\label{can rep unique lemma}
{\rm The canonical weight sequence is a valid representation for rooted trees. 

\textit{Proof} Let $R_1$ and $R_2$ be rooted trees such that ${\bf cws}(R_1) = {\bf cws}(R_2)$. Let $\tilde{R}_1$ and $\tilde{R}_2$ be canonically ordered trees that are $r$-isomorphic to $R_1$ and $R_2$, respectively. Then
\[
{\bf ws}(\tilde{R}_1) = {\bf cws}(R_1) = {\bf cws}(R_2) = {\bf ws}(\tilde{R}_2).
\]
So $\tilde{R}_1$ and $\tilde{R}_2$ are $o$-isomorphic by Lemma \ref{ord rep lemma}, and thus $r$-isomorphic. Therefore $R_1$ and $R_2$ are $r$-isomorphic.
\hfill\eot}
\end{lemma}
It immediately follows from this result that, subject to labelling, we may represent any rooted tree by a unique canonically ordered tree.

It is straightforward to show that the ordered tree $R_{max}$ that has the lexicographically largest weight sequence of all ordered trees $r$-isomorphic to $R$ is canonically ordered, and that ${\bf cws}(R) = {\bf ws}(R_{max})$.
\vspace{-2.5mm}
\subsection{\normalsize Free trees}\label{free trees}\vspace{-2.5mm}
We extend the definition of a valid representation by integer sequences to free trees: a {\em valid} representation for free trees is a well-defined representation such that any two free trees that have the same representation sequence are $f$-isomorphic.

Now, since the canonical weight sequence is a valid representation for rooted trees by Lemma \ref{can rep unique lemma}, two $f$-isomorphic rooted trees that are not $r$-isomorphic must have different canonical weight sequences. So, in order to define a valid representation for free trees using weight sequences, we need to choose a unique representative from each $f$-isomorphism class of rooted trees.

Let $T$ be a free tree of order $n$. Most algorithms for generating free trees of a given order choose the root of $T$ to be a {\em central} vertex ($T$ contains either a single central vertex or two adjacent central vertices). Instead, in keeping with our choice of the use of the weight sequence rather than the level or parent sequence, we choose the root of $T$ to be the {\em centroid} when $T$ is {\em unicentroidal}; when $T$ is {\em bicentroidal}, we represent $T$ as an ordered pair of subtrees rooted at the two centroidal vertices.
 
A {\em centroidal} vertex $u$ of $T$ is a vertex such that each component of the forest $T-u$ is of order at most $\frac{n}{2}$. It is well known that a tree is either unicentroidal, having a single centroidal vertex (in which case the largest component of $T-u$ is of order at most $\frac{n-1}{2}$), or bicentroidal, having two adjacent centroidal vertices (in which case the largest component of $T-u$ is of order $\frac{n}{2}$); see \cite{b&m1}. Moreover, it is easy to show that the centroids of two $f$-isomorphic free trees must map to each other under any $f$-isomorphism. We therefore consider the two types of free tree separately.

Suppose first that $T$ is unicentroidal. We now define the {\em free weight sequence} ${\bf fws}(T)$ of $T$ to be the canonical weight sequence of any tree $R$ that is rooted at its centroid and is $f$-isomorphic to $T$; so ${\bf fws}(T) = {\bf cws}(R)$. We note that, since the centroid consists of a single vertex and the canonical weight sequence is well defined, the free weight sequence is well defined for all unicentroidal trees. It immediately follows from Lemma \ref{can rep unique lemma} that, subject to labelling, we may represent any unicentroidal tree by a unique canonically ordered tree {\em rooted at its centroid}.

For example, suppose that the tree in Figure $1$ is a free tree $T$ (so not rooted). It is easy to see that $v_2$ is the unique centroidal vertex of $T$, and therefore $T$ is $f$-isomorphic to the canonically ordered tree in Figure $3$, which is rooted at its centroid $u$. Therefore ${\bf fws}(T) = 10\,\, 4\,2\,1\,1\, 4\,1\,1\,1\,1$.

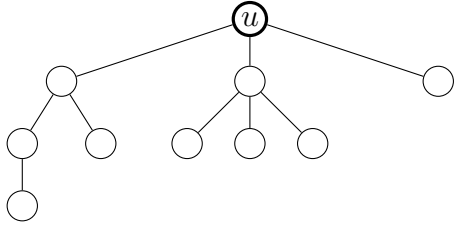
\begin{figure}[t]
\begin{centering}
\begin{tikzpicture}[sibling distance=2em, level distance = 2em, 
  every node/.style = {shape=circle, draw,  inner sep=0pt, text width=4mm, align=center}]
  \node [very thick]{$u$}
    child [sibling distance=6em] { node {} 
    child  [sibling distance=2.5em]{ node {}  
    child { node {} }  } 
    child  [sibling distance=2.5em] { node {} } }
      child { node {}
        child { node {} }
        child { node {} }
        child { node {} } }
      child  [sibling distance=6em] { node {} } ;
\end{tikzpicture}
\caption{Unicentroidal free tree representation of the tree in Figure \ref{fig 1}}\label{fig 3}
\end{centering}
\end{figure}
\newpage

\begin{lemma}\label{unicentroidal trees}
{\rm Let $T$ and $T^\prime$ be two unicentroidal free trees. If ${\bf fws}(T) = {\bf fws}(T^\prime)$ then $T$ is $f$-isomorphic to $T^\prime$.

\textit{Proof} Let $R$ and $R^\prime$ be two rooted trees, rooted at their centroids, that are $f$-isomorphic to $T$ and $T^\prime$, respectively. Suppose that ${\bf fws}(T)={\bf fws}(T^\prime)$. Then ${\bf cws}(R) = {\bf cws}(R^\prime)$, so $R$ is $r$-isomorphic to $R^\prime$ by Lemma \ref{can rep unique lemma}. Hence $T$ is $f$-isomorphic to $T^\prime$.
\hfill\eot}
\end{lemma}

We now consider the case when $T$ is bicentroidal with centroidal vertices $u$ and $v$. If we delete the edge between $u$ and $v$, we obtain disjoint trees $T_u$ and $T_v$ of order $\f{n}{2}$, which we may consider to be rooted at $u$ and $v$, respectively. We may therefore represent $T$ as the ordered pair $<\!\!T_u,\, T_v\!\!>$ when ${\bf cws}(T_u) \ge {\bf cws}(T_v)$, or $<\!\!T_v,\, T_u\!\!>$ when ${\bf cws}(T_v) \ge {\bf cws}(T_u)$. We define ${\bf fws}(T)$, the {\em free weight sequence} of $T$, to be ${\bf cws}(T_u) \smalloplus {\bf cws}(T_v)$ in the former case, and ${\bf cws}(T_v) \smalloplus {\bf cws}(T_u)$ in the latter case. We note that the first and $\frac{n+2}{2}$th elements of ${\bf fws}(T)$ correspond to $u$ and $v$, and are both equal to $\frac{n}{2}$. 

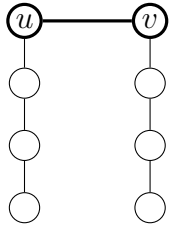
\begin{figure}[h]
\begin{centering}
\begin{tikzpicture}[level distance = 2em, 
  every node/.style = {shape=circle, draw, inner sep =0, text width=4mm, align=center}]
\node [very thick]{$u$}
child[grow=south] { node {} 
child{ node {} 
child { node {}  } } }
child[very thick, level distance = 4em, grow=east, text width=4mm, align=center, very thick] {node{$v$}
child[thin,level distance = 2em, grow=south] {node {} 
child{ node {}
child { node {} } }} };
\end{tikzpicture}
\caption{Representation of $P_8$ as a ordered pair of canonically ordered trees of order $4$}\label{fig 4}
\end{centering}
\end{figure}

Since the canonical weight sequence is well defined for rooted trees, it follows that the free weight sequence is well defined for bicentroidal trees. It immediately follows from Lemma \ref{can rep unique lemma}, that, subject to labelling, we may represent any bicentroidal tree of order $n$ by a unique ordered pair of canonically ordered trees of order $\frac{n}{2}$ (not generally rooted at their centroids). For example, the path $P_8$ is $f$-isomorphic to the tree in Figure \ref{fig 4} with centroidal vertices $u$ and $v$. Therefore ${\bf fws}(P_8) = 4\,3\,2\,1\, 4\,3\,2\,1$.

\begin{lemma}\label{bicentroidal trees}
{\rm Let $T$ and $T^\prime$ be two bicentroidal free trees. If ${\bf fws}(T)={\bf fws}(T)$ then $T$ is $f$-isomorphic to $T^\prime$.

\textit{Proof} Let $\{u,\, v\}$ and $\{u^\prime,\, v^\prime\}$ be the centroidal vertices of $T$ and $T^\prime$, respectively, and let  $<\!\!T_u,\, T_v\!\!>$ and $<\!\!T^\prime_{u^\prime},\, T^\prime_{v^\prime}\!\!>$ be the representations of $T$ and  $T^\prime$, respectively, described above. Suppose that ${\bf fws}(T)={\bf fws}(T^\prime)$. Then ${\bf cws}(T_u) = {\bf cws}(T^\prime_{u^\prime})$ and ${\bf cws}(T_v) = {\bf cws}(T^\prime_{v^\prime})$. So, by Lemma \ref{can rep unique lemma}, $T_u$ and $T_v$ are $r$-isomorphic to $T^\prime_{u^\prime}$ and $T^\prime_{v^\prime}$, respectively. Since we can recover $T$ from $T_u$ and $T_v$ by adding an edge between $u$ and $v$,  and similarly for $T^\prime$, it immediately follows that $T$ is $f$-isomorphic to $T^\prime$.
\hfill\eot}
\end{lemma}
\newpage
\begin{lemma}\label{free trees valid rep lemma}
{\rm The free weight sequence is a valid representation for free trees. 

\textit{Proof} If two free trees are isomorphic then they are both unicentroidal or both bicentroidal. The result then follows from Lemmas \ref{unicentroidal trees} and \ref{bicentroidal trees}.
\hfill\eot}
\end{lemma}
\vspace{-2.5mm}
\section{\normalsize Rooted tree generation}\label{rooted tree generation}\vspace{-2.5mm}
By Lemma \ref{can rep unique lemma}, the canonical weight sequence is a valid representation for rooted trees. So, to generate all rooted trees of order $n$, we only need to generate every possible canonical weight sequence of length $n$.

An ordered set of integer sequences $[{\bf a}_1,\, {\bf a}_2, \ldots , {\bf a}_p]$ is said to be {\em reverse lexicographically} ({\em relex}) {\em ordered} if ${\bf a}_i \ge {\bf a}_j$ when $i < j$, for all $i$ and $j$. Let $\mathcal B(n)$ denote the relex ordered set of the canonical weight sequences of all rooted trees of order $n$. It follows from Lemmas \ref{can rep unique lemma} and \ref{R_k  lemma 1} that, for each element ${\bf s} = s_1\, s_2\, \ldots s_n$ of $\mathcal B(n)$, there exists a unique canonically ordered tree $R$, with vertices labelled $v_1,\, v_2, \ldots ,v_n$ in pre-order, such that ${\bf ws}(R) =  {\bf s}$, where $s_k = wt(v_k)$ for all $k$, $1 \le k \le n$.

If ${\bf s} = s_1\, s_2\, \ldots s_n$ is an integer sequence, we let ${\bf s^{\#}} = s_2\, s_3\, \ldots s_n$, i.e., ${\bf s^{\#}}$ is ${\bf s}$ with the first element $s_1$ removed. So if ${\bf s}$ is the weight sequence of an ordered tree $R$, then ${\bf s^{\#}}$ is the weight sequence of the ordered {\em forest} obtained by removing the root of $R$.

We write ${\bf s} \succcurlyeq {\bf t}$ if ${\bf t}$ is some other integer sequence such that either ${\bf s} \ge {\bf t}$ or ${\bf s}$ is a prefix of ${\bf t}$, i.e.,  ${\bf t} = {\bf s} \smalloplus {\bf x}$ for some integer sequence {\bf x}.

Let $\mathcal A_q(n)$ be the set of all ordered pairs $<\!\!{\bf a},\, {\bf b}\!\!>$ in $\mathcal B(q) \times \mathcal B(n-q)$ such that ${\bf a} \succcurlyeq \bhash$, and let $\mathcal A(n, \, maxq) = \bigcup\limits_{q=1}^{maxq} \mathcal A_q(n)$. We recall that if $<\!\!{\bf a},\, {\bf b}\!\!>$ is in $\mathcal A_q(n)$ then the first element of {\bf a} is $q$, $|{\bf a}| = q$, the first element of {\bf b} is $n-q$ and $|{\bf b}| = n-q$.
\begin{lemma}\label{B(n) lemma 1}
{\rm There is a bijection $\beta$ from $\mathcal A(n,\, n-1)$ to $\mathcal B(n)$ defined by
\begin{equation}
\beta(<\!\!{\bf a},\, {\bf b}\!\!>) = n \smalloplus {\bf a} \smalloplus  \bhash. \label{B(n) lemma 1 eqn}
\end{equation}
\textit{Proof} Suppose that $<\!\!{\bf a},\, {\bf b}\!\!>$ is in $\mathcal A_q(n)$, for some $q$, $1 \le q \le n-1$. We first show that $\beta(<\!\!{\bf a},\, {\bf b}\!\!>) \in \mathcal B(n)$.
 
Let $R_1$ be a canonically ordered tree rooted at $v$ such that ${\bf ws}(R_1) = {\bf a}$, and let $R_2$ be a canonically ordered tree rooted at $u$ such that ${\bf ws}(R_2) = {\bf b}$. Let $u_1, \,u_2, \ldots, u_p$ be the children of $u$ in order, and let $R$ be a new ordered tree rooted at $u$ with children $v,\, u_1, \,u_2, \ldots, u_p$, i.e., $R$ is obtained from $R_2$ by adding $R_1$ as the new first subtree of $u$. Now $<\!\!{\bf a},\, {\bf b}\!\!>$ is in $\mathcal A_q(n)$, so ${\bf a} \succcurlyeq {\bf b^{\#}}$, and thus $wt(v) \ge wt(u_1)$. Therefore $R$ is canonically ordered as both $R_1$ and $R_2$ are canonically ordered. So ${\bf ws}(R)$ is in $\mathcal B(n)$ and, moreover, ${\bf ws}(R) = n  \smalloplus {\bf a}  \smalloplus \bhash$ by Corollary \ref{R_k  cor 2}. Therefore $\beta(<\!\!{\bf a},\, {\bf b}\!\!>) \in \mathcal B(n)$.
 
Suppose that $<\!\!{\bf a}_0,\, {\bf b}_0\!\!>$ is in $\mathcal A_r(n)$, for some $r$, and that $ n \smalloplus {\bf a}_0 \smalloplus {\bf b}_0^{\#} = n \smalloplus {\bf a} \smalloplus \bhash$. Then $r=q$ since the first element of ${\bf a}_0$ must be equal to the first element of ${\bf a}$. It follows that ${\bf a}_0= {\bf a}$ and ${\bf b}_0= {\bf b}$, as $|{\bf a}_0|=|{\bf a}|$. Hence $\beta$ is injective.

Now suppose that ${\bf s} = s_1\,  s_2 \, \ldots s_n$ is an element of $\mathcal B(n)$, and let $R$ be the unique canonically ordered tree such that ${\bf ws}(R) = {\bf s}$. By Lemma \ref{R_k  lemma 1}(b) and (c), ${\bf ws}(R(v_2)) =  s_2\,  s_3 \, \ldots s_{s_{2} +1}$ and ${\bf ws}(R - R(v_2)) = t\,  s_{{s_{2} + 2}} \,  s_{{s_{2}+3}} \ldots s_n$ where $t = n-s_2$. Clearly, since $R$ is canonically ordered, so are $R(v_2)$ and $R - R(v_2)$. Hence ${\bf ws}(R(v_2)) \in \mathcal B(s_2)$ and ${\bf ws}(R - R(v_2)) \in \mathcal B (n - s_2)$. Moreover, since $R$ is canonically ordered, it follows from Corollary \ref{R_k  cor 2} and the definition of $\succcurlyeq$ that $s_2\,  s_3 \, \ldots s_{s_{2} +1} \succcurlyeq s_{{s_{2} + 2}} \,  s_{{s_{2}+3}} \ldots s_n$. Therefore $<\!\! {\bf ws}(R(v_2)), \, {\bf ws}(R - R(v_2))\!\!>$ is in $\mathcal A_{s_2}(n)$. Hence $\beta$ is onto, and is therefore a bijection.
\hfill\eot}
\end{lemma}
\begin{cor}\label{B(n) cor 2}
{\rm For any $n$, $\mathcal B(n)$ can be constructed from the sets $\mathcal B(q)$, where $1 \le q \le n - 1$.

\textit{Proof} It is easy to construct all rooted trees, and therefore $\mathcal B(n)$, when $n \le 3$. The result then follows using equation (\ref{B(n) lemma 1 eqn}) and induction on $n$.
\hfill\eot}
\end{cor}
The image of $\mathcal A_q(n)$ under the bijection $\beta$ defined in (\ref{B(n) lemma 1 eqn}) is denoted by $\mathcal B_q(n)$, i.e., $\mathcal B_q(n)$ corresponds to those rooted trees of order $n$ for which the first subtree of the root is of order $q$. So $\mathcal B_q(n)$ contains those sequences in $\mathcal B(n)$ for which the second element is $q$. Clearly $\mathcal B(n) = \bigcup\limits_{q=1}^{n-1} \mathcal B_q(n)$. 

Following along the lines of the proofs of Lemma \ref{B(n) lemma 1} and Corollary \ref{B(n) cor 2}, we now construct a simple recursive algorithm to generate the elements of $\mathcal B(n)$. For each $q$, $1 \le q \le n-1$, and for each ${\bf a}$ in $\mathcal B(q)$, we need to find those elements ${\bf b}$ in $\mathcal B(n - q)$ for which $\bf a \succcurlyeq \bhash$. We then form the integer sequence $n \smalloplus {\bf a} \smalloplus \bhash$ to obtain the appropriate element of $\mathcal B(n)$. We can avoid searching the whole of $\mathcal B(n - q)$ for those elements ${\bf b}$ for which $\bf a \succcurlyeq \bhash$, by noting that we only need to consider those elements that are in $\mathcal B_r(n-q)$, where $1 \le r \le \min(n-q-1, q)$.

In the pseudocode we use in the rest of the paper, we represent lists in square brackets; we use $\oplus$ for concatenating lists, and continue to use $\smalloplus$ for concatenating integer sequences. If L is a list, then L[{\em start} ...] denotes the sublist beginning at element L[{\em start}] and ending at the last element of L.

The following function \smallsf{RootedTrees}(n) generates $\mathcal B(n)$. It makes use of the helper function \\ \smallsf{RTHelper}($n, q$) that generates $\mathcal B_q(n)$.
\begin{algorithm}
\SetDataSty{smallsf}
\SetFuncSty{smallsf}
\SetKwSty{bfsf}
\SetKw{Downto}{downto}
\SetKw{From}{from}
\SetKwProg{Function}{Function}{}{}
\SetKwFunction{RootedTrees}{RootedTrees}
\SetKwFunction{RTHelper}{RTHelper}
\SetKwData{Bn}{Bn}
\SetAlgoNoEnd\SetAlgoNoLine\DontPrintSemicolon
\setstretch{1.1}
\Function{\RootedTrees{n}}
{
\lIf{$n = 1$}{\Return $[1]$}
\Bn \la [ ]\;
\For{$q$ \From $n - 1$ \Downto $1$}
	{\Bn \la \Bn $\oplus$ \RTHelper{$n, q$}\;}
{\Return \Bn}
}
\end{algorithm}

\begin{algorithm}
\SetDataSty{smallsf}
\SetFuncSty{smallsf}
\SetKwSty{bfsf}
\SetKw{Downto}{downto}
\SetKw{From}{from}
\SetKw{In}{in}
\SetKwProg{Function}{Function}{}{}
\SetKwFunction{RootedTrees}{RootedTrees}
\SetKwFunction{RTHelper}{RTHelper}
\SetKwData{Bqn}{Bqn}
\SetKwData{a}{a}
\SetKwData{b}{b}
\SetAlgoNoEnd\SetAlgoNoLine\DontPrintSemicolon
\setstretch{1.1}
\Function{\RTHelper{$n, q$}}
{
\Bqn \la [ ]\;
$newq$ \la $\min(n - q - 1, q)$\;
{\uIf{$newq = 0$}
	{\lFor{\a \In \RootedTrees{$q$}}{\Bqn \la \Bqn $\oplus$ [$n \, \smalloplus$ \a]}}
\Else{
	\For{\a \In \RootedTrees{$q$}}
		{\For{$r$ \From $newq$ \Downto $1$}
			{\For{\b \In \RTHelper{$n-q, r$}}
				{\lIf{\a $\succcurlyeq$  \b$^{\!\!\#}$}{\Bqn \la  \Bqn $\oplus$ [$n\, \smalloplus$ {\a} $\smalloplus$ \b$^{\!\!\#}$]}}
			}
		}
	}
	}
{\Return \Bqn}
}
\end{algorithm}

There are two key points to note about the recursive calls in \smallsf{RTHelper}. Firstly, the length of the subsequence corresponding to the first subtree of the root must be smaller than the order of the tree itself; so we always have $q < n$. Secondly, if $q = n-1$ then the sequence represents a tree in which the root has only one subtree; so we simply return $n$ concatenated with the subsequence that corresponds to this subtree.

We note that $\mathcal B_r(n-q)$, the list returned by \smallsf{RTHelper}($n - q, r$), will clearly require too much space for most values of $r$ when $n-q$ is large. This problem is addressed by returning a generator instead of a list (see Section \ref{generators}). We note further that, in the loops in \smallsf{RootedTrees} and \smallsf{RTHelper}, the variables $q$ and $r$ are counting down, so \smallsf{Bn} and \smallsf{Bqn} will be relex ordered, as required.

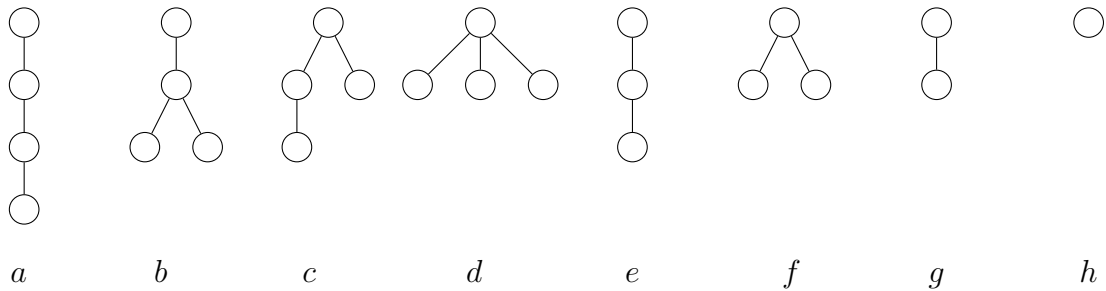
\begin{figure}[h]
\begin{centering}
\begin{tikzpicture}[sibling distance=2em, level distance = 2em,
  every node/.style = {shape=circle, draw, align=center}]
\begin{scope}[shift={(-5, 0)}]
\node {}
    child { node {}
    child { node {}
    child { node {}}}};
\end{scope}
\begin{scope}[shift={(-3, 0)}]
\node {}
    child { node {}
    child { node {} }
    child { node {}}};
\end{scope}
\begin{scope}[shift={(-1, 0)}]
\node {}
    child { node {}
    child { node {} }}
    child { node {}};
\end{scope}
\begin{scope}[shift={(1, 0)}]
\node {}
    child { node {}}
    child { node {}}
   child { node {}};
\end{scope}
\begin{scope}[shift={(3, 0)}]
\node {}
    child { node {}
    child { node {}}};
\end{scope}
\begin{scope}[shift={(5, 0)}]
\node {}
    child { node {}}
    child { node {}};
\end{scope}
\begin{scope}[shift={(7, 0)}]
\node  {}
child {node {}};
\end{scope}
\begin{scope}[shift={(9, 0)}]
  \node {};
\end{scope}
\end{tikzpicture}

\vspace{2.5mm}
\hspace{-2mm} $a$ \hspace{14mm} $b$ \hspace{15mm} $c$ \hspace{17mm} $d$ \hspace{16mm} $e$ \hspace{16mm} $f$ \hspace{14mm} $g$ \hspace{15mm} $h$

\vspace{2.5mm}
\caption{All canonically ordered trees of order at most 4}\label{rooted trees order at most 4}
\end{centering}
\end{figure}

For example, consider the sets $\mathcal B(4)=[4\, 3\, 2\, 1,\, 4\, 3\, 1\, 1, \,4\, 2\, 1\, 1, \,4\, 1\, 1\, 1]$,  $\mathcal B(3)=[3\, 2\, 1,\,3\, 1\, 1]$,  $\mathcal B(2)=[2\, 1]$ and  $\mathcal B(1)=[1]$, which correspond to the canonically ordered trees in Figure \ref{rooted trees order at most 4} (we have omitted the vertex labels for simplicity).

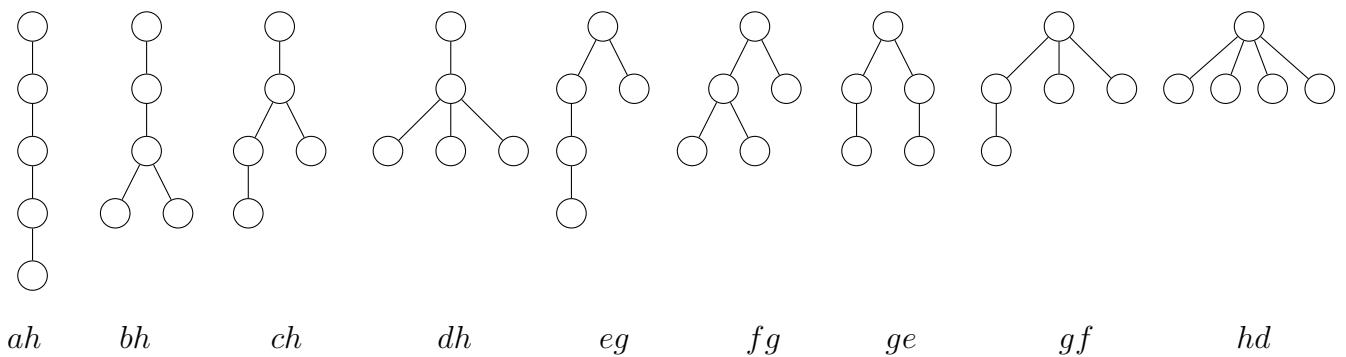
\begin{figure}[h]
\begin{centering}
\begin{tikzpicture}[sibling distance=2em, level distance = 2em, 
  every node/.style = {shape=circle, draw, align=center}]
\begin{scope}[shift={(-5, 0)}]
\node {}
    child { node {}
    child { node {}
    child { node {}
    child { node {}}}}};
\end{scope}
\begin{scope}[shift={(-3.5, 0)}]
\node {}
    child { node {}
    child { node {}
    child { node {} } 
    child { node {}}} };
\end{scope}
\begin{scope}[shift={(-1.75, 0)}]
\node {}
 child {node {}
    child { node {}
    child { node {} } }
    child { node {} } };
\end{scope}
\begin{scope}[shift={(0.5, 0)}]
\node {}
 child {node {} 
    child { node {} } 
    child { node {} }  
    child { node {} } } ;
\end{scope}
\begin{scope}[shift={(2.5, 0)}]
\node {}
    child { node {}
    child { node {}
    child { node {}}}}
    child { node {}};
\end{scope}
\begin{scope}[shift={(4.5, 0)}]
\node {}
    child { node {}
    child { node {} }
    child { node {}}}
    child { node {}};
\end{scope}
\begin{scope}[shift={(6.25, 0)}]
\node {}
    child { node {}
    child { node {} } }
    child { node {}
    child { node {} } };
\end{scope}
\begin{scope}[shift={(8.5, 0)}]
\node {}
 child {node {} 
    child { node {} } }
 child { node {} } 
child { node {} };
\end{scope}
\begin{scope}[shift={(11, 0)}]
\node {}[sibling distance=1.5em]
 child {node {} }
    child { node {} } 
 child { node {} } 
child { node {} };
\end{scope}
\end{tikzpicture}

\vspace{2.5mm}
\hspace{-7mm} $ah$ \hspace{7.5mm} $bh$ \hspace{13mm} $ch$ \hspace{15mm} $dh$ \hspace{14mm} $eg$ \hspace{12.5mm} $fg$ \hspace{11mm} $ge$ \hspace{16mm} $gf$ \hspace{16mm} $hd$ 
\caption{Canonically ordered trees of order $5$}\label{rooted trees order 5}
\end{centering}
\end{figure}

Recalling that $\mathcal B_q(n)$ contains those sequences in $\mathcal B(n)$ for which the second element is $q$, it is straightforward to show that the call \smallsf{RootedTrees}(5) yields the set
\[
\mathcal B(5)= [5\, 4\, 3\, 2\, 1,\, 5\, 4\, 3\, 1\, 1, \,5\, 4\, 2\, 1\, 1, \,5\, 4\, 1\, 1\, 1,\, 5\, 3\, 2\, 1\, 1,\,5\, 3\, 1\, 1\, 1,\, 5\, 2\, 1\, 2\, 1,\, 5\, 2\, 1\, 1\, 1,\, 5\, 1\, 1\, 1\, 1].
\]
These correspond to the canonically ordered trees in Figure \ref{rooted trees order 5}, where the label indicates which pair of trees in Figure \ref{rooted trees order at most 4} - corresponding to {\bf a} and {\bf b} in equation (\ref{B(n) lemma 1 eqn}) - are used to construct the tree.

We discuss some optimisations of the function \smallsf{RTHelper} in Section \ref{code efficiencies}.

\newpage
\section{\normalsize Free tree generation}\label{free tree generation}\vspace{-2.5mm}
By Lemma \ref{free trees valid rep lemma}, the free weight sequence is a valid representation for free trees. So, in order to generate all free trees of order $n$, we only need to generate every possible free weight sequence of length $n$. We recall that a free tree is either unicentroidal or bicentroidal, which have slightly different definitions of the free weight sequence.

We denote the relex ordered set of the free weight sequence representations of all free trees, unicentroidal free trees and bicentroidal free trees of order $n$ by $\mathcal F(n)$, $\mathcal F_U(n)$ and $\mathcal F_B(n)$, respectively. So $\mathcal F(n) = \mathcal F_U(n) \oplus \mathcal F_B(n)$, i.e., the elements of $\mathcal F_U(n)$ followed by those of $\mathcal F_B(n)$.
\vspace{-2.5mm}
\subsection{\normalsize Unicentroidal}\label{unicentroidal}\vspace{-2.5mm}
We recall from Section \ref{free trees} that the free weight sequence ${\bf fws}(T)$ of a unicentroidal free tree $T$ is the canonical weight sequence of any tree rooted at its centroid that is $f$-isomorphic to $T$. So \\$\mathcal F_U(n) \subseteq \mathcal B(n)$. We can therefore generate $\mathcal F_U(n)$ using a simple modification of the algorithm \smallsf{RootedTrees} from Section \ref{rooted tree generation}: the canonically ordered tree $R$ that represents a unicentroidal free tree $T$ is rooted at its centroid, so the sub-trees of the root are of order at most $\frac{n - 1}{2}$. It follows that $|{\bf a}|  \le \left\lfloor\frac{n - 1}{2} \right\rfloor$ for every pair $<\!\!{\bf a},\, {\bf b}\!\!>$ in $\mathcal A(n,\,n-1)$ for which $\beta(<\!\!{\bf a},\, {\bf b}\!\!>)$ is in $\mathcal F_U(n)$.
\begin{lemma}\label{fws lemma free trees}
{\rm The mapping $\beta$ defined in equation (\ref{B(n) lemma 1 eqn}) is a bijection from $\mathcal A(n,\,\left\lfloor\frac{n - 1}{2} \right\rfloor)$ to $\mathcal F_U(n)$.

\textit{Proof} We may represent any unicentroidal free tree $T$ by a unique canonically ordered tree $R$ in which the weight of each child of the root of $R$ is at most $\left\lfloor\frac{n - 1}{2} \right\rfloor$. So the result can be proved in a similar manner to Lemma \ref{B(n) lemma 1}, with the additional restriction that $|{\bf a}|  \le \left\lfloor\frac{n - 1}{2} \right\rfloor$, i.e., we use $\mathcal A(n,\,\left\lfloor\frac{n - 1}{2} \right\rfloor)$ instead of $\mathcal A(n, n-1)$.
\hfill\eot}
\end{lemma}
\begin{cor}\label{cor free trees}
{\rm For any $n$, $\mathcal F_U(n)$ can be constructed from the sets $\mathcal B(q)$, where $1 \le q \le n - 1$.
\hfill\eot}
\end{cor}
The following function \smallsf{UFT}(n) generates the set $\mathcal F_U(n)$.  It also makes use of the helper function \smallsf{RTHelper}($n, q$).
\begin{algorithm}
\SetDataSty{smallsf}
\SetFuncSty{smallsf}
\SetKwSty{bfsf}
\SetKw{Downto}{downto}
\SetKw{From}{from}
\SetKw{In}{in}
\SetKwProg{Function}{Function}{}{}
\SetKwFunction{UFT}{UFT}
\SetKwFunction{RTHelper}{RTHelper}
\SetKwData{UFn}{UFn}
\SetAlgoNoEnd\SetAlgoNoLine\DontPrintSemicolon
\setstretch{1.1}
\Function{\UFT{$n$}}
{\lIf{$n = 1$}{\Return $[1]$}{\UFn \la [ ]}\;
$maxq$ \la $\left\lfloor \frac{1}{2}(n-1) \right\rfloor$\;
\For{$q$ \From maxq \Downto $1$}
	{\UFn \la \UFn $\oplus$ \RTHelper{$n, q$}\;}
{\Return \UFn}
}
\end{algorithm}

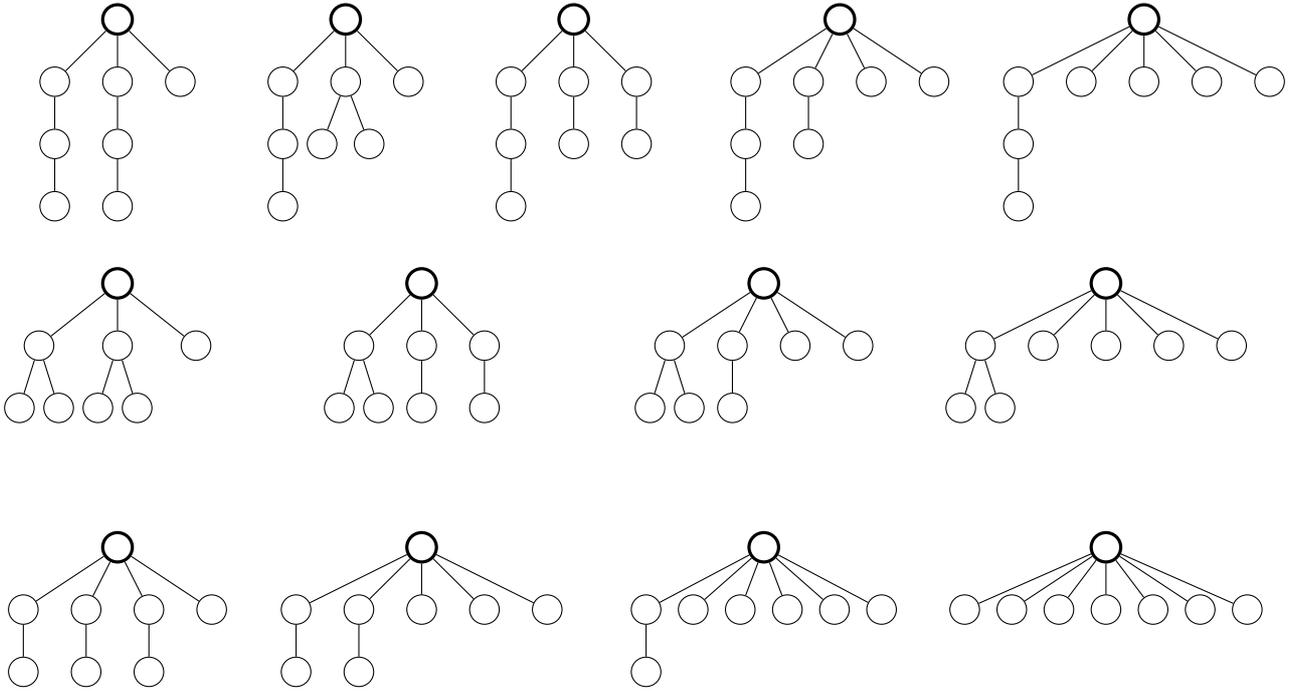
\begin{figure}[h]
\begin{tikzpicture}[sibling distance=2em, level distance = 2em, 
  every node/.style = {shape=circle, draw, align=center}]
\begin{scope}[shift={(-8, 0)}]
\node[very thick] {}
    child { node {}
    child { node {}  
    child { node {} } } }
    child { node {}
    child { node {}  
    child { node {} } } }
	child { node {} };
\end{scope}
\begin{scope}[shift={(-5, 0)}]
\node[very thick] {}
    child { node {}
    child { node {}  
    child { node {} } } }
    child { node {}
[sibling distance=1.5em]
    child { node {} }  
    child { node {} } }
	child { node {} };
\end{scope}
\begin{scope}[shift={(-2, 0)}]
\node[very thick] {}
    child { node {}
    child { node {}  
    child { node {} } } }
    child { node {}
    child { node {} } }  
    child { node {}
    child { node {} } } ;
\end{scope}
\begin{scope}[shift={(1.5, 0)}]
\node[very thick] {}
    child { node {}
    child { node {}  
    child { node {} } } }
    child { node {}
    child { node {} } }  
    child { node {} }
    child { node {} } ;
\end{scope}
\begin{scope}[shift={(5.5, 0)}]
\node[very thick] {}
    child { node {}
    child { node {}  
    child { node {} } } }
    child { node {} }
    child { node {} }  
    child { node {} }
    child { node {} } ;
\end{scope}
\begin{scope}[shift={(-8, -3.5)}]
\node[very thick] {}
[sibling distance=2.5em]
    child { node {}
[sibling distance=1.25em]
    child { node {} }  
    child { node {} } }
    child { node {}
[sibling distance=1.25em]
    child { node {} }  
    child { node {} } }
	child { node {} };
\end{scope}
\begin{scope}[shift={(-4, -3.5)}]
\node[very thick] {}
    child { node {}
[sibling distance=1.25em]
    child { node {} }  
    child { node {} } }
    child { node {}
    child { node {} } }  
    child { node {}
    child { node {} } } ;
\end{scope}
\begin{scope}[shift={(0.5, -3.5)}]
\node[very thick] {}
    child { node {}
[sibling distance=1.25em]
    child { node {} }  
    child { node {} } }
    child { node {}
    child { node {} } }  
    child { node {} }
    child { node {} } ;
\end{scope}
\begin{scope}[shift={(5, -3.5)}]
\node[very thick] {}
    child { node {}
[sibling distance=1.25em]
    child { node {} }  
    child { node {} } }
    child { node {} }
    child { node {} }  
    child { node {} }
    child { node {} } ;
\end{scope}
\begin{scope}[shift={(-8, -7)}]
\node[very thick] {}
    child { node {}
    child { node {} } } 
    child { node {}
    child { node {} } }
    child { node {}
    child { node {} } }
    child { node {} } ;
\end{scope}
\begin{scope}[shift={(-4, -7)}]
\node[very thick] {}
    child { node {}
    child { node {} } } 
    child { node {}
    child { node {} } }
    child { node {} }
    child { node {} } 
    child { node {} } ;
\end{scope}
\begin{scope}[shift={(0.5, -7)}]
\node[very thick] {}
[sibling distance=1.5em] 
   child { node {}
    child { node {} } } 
    child { node {} }
    child { node {} }
    child { node {} }
    child { node {} } 
    child { node {} } ;
\end{scope}
\begin{scope}[shift={(5, -7)}]
\node[very thick] {}
[sibling distance=1.5em]
    child { node {} }
    child { node {} } 
    child { node {} }
    child { node {} } 
    child { node {} }
    child { node {} } 
    child { node {} } ;
\end{scope}
\end{tikzpicture}
\caption{Canonically ordered trees corresponding to unicentroidal free trees of order $8$}\label{free uni order 8}
\end{figure}

For example, we can construct $\mathcal F_U(8)$ using the call \smallsf{UFT}(8) to obtain
\begin{eqnarray}
\mathcal F_U(8)=&[&\! \! \! \! \! \!  8 \, 3\, 2\, 1\, 3\, 2\, 1\, 1,\,8 \, 3\, 2\, 1\, 3\, 1\, 1\, 1,\,8 \, 3\, 2\, 1\, 2\, 1\, 2\, 1,\,8 \, 3\, 2\, 1\, 2\, 1\, 1\, 1, 8 \, 3\, 2\, 1\, 1\, 1\, 1\, 1] \nonumber\\
&\oplus&[8 \, 3\, 1\, 1\, 3\, 1\, 1\, 1,\,8 \, 3\, 1\, 1\, 2\, 1\, 2\, 1,\,8 \, 3\, 1\, 1\, 2\, 1\, 1\, 1,\,8 \, 3\, 1\, 1\, 1\, 1\, 1\, 1] \nonumber\\
&\oplus& [8 \, 2\, 1\, 2\, 1\, 2\, 1\, 1,\,8 \, 2\, 1\, 2\, 1\, 1\, 1\, 1,\,8 \, 2\, 1\, 1\, 1\, 1\, 1\, 1,\,8 \, 1\, 1\, 1\, 1\, 1\, 1\, 1].\nonumber
\end{eqnarray}
This corresponds to the set of canonically ordered trees in Figure \ref{free uni order 8}. 
\vspace{2.5mm}
\vspace{-5mm}
\newpage
\subsection{\normalsize Bicentroidal}\label{bicentroidal}\vspace{-2.5mm}
We recall from Section \ref{free trees} that the free weight sequence of a bicentroidal free tree with centroidal vertices $u$ and $v$ is ${\bf cws}(T_u) \smalloplus {\bf cws}(T_v)$, where ${\bf cws}(T_u) \ge {\bf cws}(T_v)$. 
\begin{lemma}\label{wt lemma free trees}
{\rm For any $n$, $\mathcal F_B(n)$ can be constructed from the set $\mathcal B(\frac{n}{2})$.
\hfill\eot}
\end{lemma}

The following function \smallsf{BFT}(n) generates the set $\mathcal F_B(n)$.  
\begin{algorithm}
\SetDataSty{smallsf}
\SetFuncSty{smallsf}
\SetKwSty{bfsf}
\SetKw{Downto}{downto}
\SetKw{From}{from}
\SetKw{In}{in}
\SetKwProg{Function}{Function}{}{}
\SetKwFunction{BFT}{BFT}
\SetKwFunction{RootedTrees}{RootedTrees}
\SetKwData{BFn}{BFn}
\SetKwData{a}{a1}
\SetKwData{aa}{a2}
\SetAlgoNoEnd\SetAlgoNoLine\DontPrintSemicolon
\setstretch{1.1}
\Function{\BFT{n}}
{
\BFn \la [ ]\;
\For{\a \In \RootedTrees{$\frac{n}{2}$}}
	{\For{\aa \In \RootedTrees{$\frac{n}{2}$}}
		{\lIf{\a $\ge$ \aa}{\BFn \la \BFn $\oplus$ [\a $\smalloplus$ \aa]}}
	}
{\Return \BFn}
}
\end{algorithm}

For example, we can construct $\mathcal F_B(8)$ using the call {\smallsf{BFT}(8) to obtain
\begin{eqnarray}
\mathcal F_B(8)=&[&\! \! \! \! \! \! 4 \, 3\, 2\, 1\, 4\, 3\, 2\, 1,\,4 \, 3\, 2\, 1\, 4\, 3\, 1\, 1,\,4 \, 3\, 2\, 1\, 4\, 2\, 1\, 1,\,4 \, 3\, 2\, 1\, 4\, 1\, 1\, 1] \nonumber\\
&\oplus&[4 \, 3\, 1\, 1\, 4\, 3\, 1\, 1,\,4 \, 3\,  1\, 1\, 4\, 2\, 1\, 1,\,4 \, 3\, 1\, 1\, 4\, 1\, 1\, 1]\\ \nonumber
&\oplus&[4 \, 2\, 1\, 1\, 4\, 2\, 1\, 1,\,4 \, 2\, 1\, 1\, 4\, 1\, 1\, 1,\,4 \, 1\, 1\, 1\, 4\, 1\, 1\, 1] \nonumber
\end{eqnarray}
This corresponds to the set of ordered pairs of canonically ordered rooted trees of order $4$ (see Figure \ref{rooted trees order at most 4}), with an additional edge joining their roots, as shown in Figure \ref{free bi order 8}.

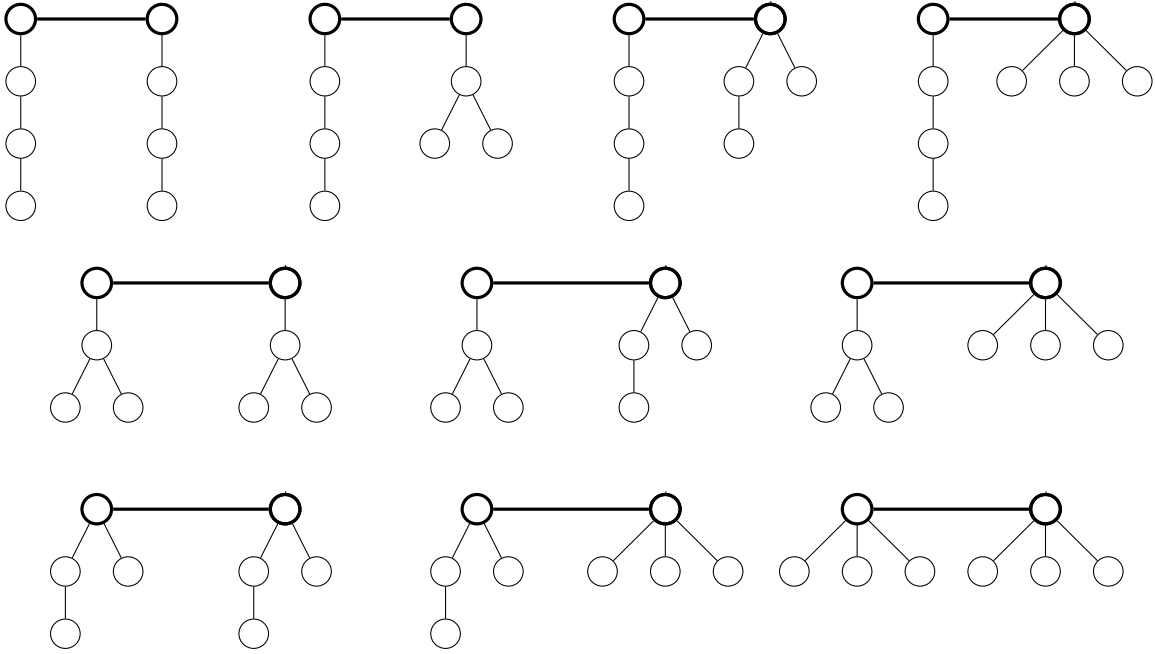
\begin{figure}[h]
\centering
\begin{tikzpicture}[sibling distance=2em, level distance = 2em, 
  every node/.style = {shape=circle, draw, align=center}]
\begin{scope}[shift={(-5, 0)}]
\node [very thick]{}
child[grow=south] { node {} 
child { node {} 
child { node {}  } } }
child[very thick, level distance=4.5em, grow=east] {node{}
child[thin, level distance = 2em, grow=south] {node {} 
child{ node {}
child { node {} } }} };
\end{scope}
\begin{scope}[shift={(-1, 0)}]
\node [very thick]{}
child[grow=south] { node {} 
child { node {} 
child { node {}  } } }
child[very thick, level distance=4.5em, grow=east] {node{}
child[thin, level distance = 2em, grow=south] {node {} 
child { node {} }
child {node {} } }};
\end{scope}
\begin{scope}[shift={(3, 0)}]
\node [very thick]{}
child[grow=south] { node {} 
child{ node {} 
child { node {}  } } }
child[very thick, level distance=4.5em, grow=east] {node{}
child[level distance=0em, grow=south]{node{}
child[level distance=2em, thin]{node {}
child { node {} }}
child[level distance=2em, thin] {node {} }}};
\end{scope}
\begin{scope}[shift={(7, 0)}]
\node [very thick]{}
child[grow=south] { node {} 
child { node {} 
child { node {}  } } }
child[very thick, level distance=4.5em, grow=east] {node{}
child[level distance=0em, grow=south]{node{}
child[thin, level distance = 2em] {node {} }
child[thin, level distance = 2em] { node {} }
child[thin, level distance = 2em] { node {} } }};
\end{scope}
\begin{scope}[shift={(-4, -3.5)}]
\node [very thick]{}
child[grow=south, level distance = 2em] {node {} 
child { node {} }
child{ node {} } }
child[very thick, level distance=6em, grow=east] {node{}
child[level distance=0em, grow=south]{node{}
child[thin, level distance = 2em] {node {} 
child { node {} }
child{ node {} } }}};
\end{scope}
\begin{scope}[shift={(1, -3.5)}]
\node [very thick]{}
child[grow=south, level distance = 2em] {node {} 
child { node {} }
child{ node {} } }
child[very thick, level distance=6em, grow=east] {node{}
child[level distance=0em, grow=south]{node{}
child[level distance=2em, thin]{node {}
child { node {} }}
child[level distance=2em, thin] {node {} }}};
\end{scope}
\begin{scope}[shift={(6,  -3.5)}]
\node [very thick]{}
child[grow=south, level distance = 2em] {node {} 
child { node {} }
child{ node {} } }
child[very thick, level distance=6em, grow=east] {node{}
child[level distance=0em, grow=south]{node{}
child[thin, level distance = 2em] {node {} }
child[thin, level distance = 2em] { node {} }
child[thin, level distance = 2em] { node {} } }};
\end{scope}
\begin{scope}[shift={(-4,-6.5)}]
\node [very thick]{}
child[level distance=0em, grow=south]{node{}
child[level distance=2em, thin]{node {}
child { node {} }}
child[level distance=2em, thin] {node {} }}
child[very thick, level distance = 6em, grow=east] {node{}
child[level distance=0em, grow=south]{node{}
child[level distance=2em, thin]{node {}
child { node {} }}
child[level distance=2em, thin] {node {} }}};
\end{scope}
\begin{scope}[shift={(1, -6.5)}]
\node [very thick]{}
child[level distance=0em, grow=south]{node{}
child[level distance=2em, thin]{node {}
child { node {} }}
child[level distance=2em, thin] {node {} }}
child[very thick, level distance = 6em, grow=east] {node{}
child[level distance=0em, grow=south]{node{}
child[thin, level distance = 2em] {node {} }
child[thin, level distance = 2em] { node {} }
child[thin, level distance = 2em] { node {} } }};
\end{scope}
\begin{scope}[shift={(6, -6.5)}]
\node [very thick]{}
child[level distance=0em, grow=south]{node{}
child[level distance = 2em, thin] {node {} }
child[level distance = 2em, thin] { node {} }
child[level distance = 2em, thin] { node {} }} 
child[very thick, level distance = 6em, grow=east] {node{}
child[level distance=0em, grow=south]{node{}
child[thin, level distance = 2em] {node {} }
child[thin, level distance = 2em] { node {} }
child[thin, level distance = 2em] { node {} } }};
\end{scope}
\end{tikzpicture}

\vspace{2.5mm}
\caption{Bicentroidal free trees of order $8$ constructed from canonically ordered pairs of rooted trees of order $4$}\label{free bi order 8}
\end{figure}
By combining the unicentroidal and bicentroidal free tree algorithms, we can generate all free trees of order $n$ using the following function \smallsf{FreeTrees}(n).
\begin{algorithm}
\SetDataSty{smallsf}
\SetFuncSty{smallsf}
\SetKwSty{bfsf}
\SetKw{Downto}{downto}
\SetKw{From}{from}
\SetKw{In}{in}
\SetKwProg{Function}{Function}{}{}
\SetKwFunction{FreeTrees}{FreeTrees}
\SetKwFunction{UFT}{UFT}
\SetKwFunction{BFT}{BFT}
\SetAlgoNoEnd\SetAlgoNoLine\DontPrintSemicolon
\setstretch{1.1}
\Function{\FreeTrees{$n$}}
{
{\Return $\UFT(n) \oplus \BFT(n)$}
}
\end{algorithm}

For example, the $23$ free trees of order $8$ is the ordered union of $\mathcal F_U(8)$ and $\mathcal F_B(8)$, i.e., the trees in Figures \ref{free uni order 8} and \ref{free bi order 8}.\vspace{-2.5mm}
\section{\normalsize Improvements and implementation of the algorithms}\label{code efficiencies}\vspace{-2.5mm}
We now outline some of the changes we have made to improve the efficiency of the functions described in Sections \ref{rooted tree generation} and \ref{free tree generation} above, and their implementations in Python. We use ${\bf a}^k$ to denote the integer sequence that is formed by the concatenation of $k$ copies of the integer sequence {\bf a}.\vspace{-2.5mm}
\subsection{\normalsize Improvements to RTHelper}\label{improvements to RTHelper}\vspace{-2.5mm}
Firstly we note that, since there is only one rooted tree of order $1$ and one of order $2$, having canonical weight sequences $1$ and $2\, 1$, respectively, we may compute the result in a more efficient and explicit manner when $q$ is $1$ or $2$. If $q = 1$ then the function should return the single sequence $n \smalloplus 1^{n-1}$, and if $q = 2$ then it should return the ordered set of sequences $(2 \, 1)^t \smalloplus 1^{n - 1 - 2t}$ for $t$ from $\left\lfloor \frac{n-1}{2} \right\rfloor$ down to $1$. We also note that, when $q = n-2$, the second subtree of the root contains just a single vertex, so {\bf b} is just $1$ in this case. These observations enable us to remove the recursive calls to \smallsf{RTHelper} when $q \in \{1,\, 2, n-2\}$, as in the more efficient function \smallsf{RTHelper2}($n, q$) below. (In practice, in the implementation, we subsume the case $q=1$ into the case $q=2$ by reducing the lower limit of $t$ from $1$ to $0$, and correspondingly increasing the lower limit of $q$ from $1$ to $2$ in \smallsf{RootedTrees}.)

Next we note that, during the execution of \smallsf{RTHelper}, checking whether {\bf a} $\succcurlyeq \bhash$ is only necessary when the order $r$ of the second child of the root is the same as the order $q$ of the first child. We further note that after  {\bf a} $\succcurlyeq \bhash$ for the first time, this will also hold for all subsequent sequences ${\bf b}$, since $\mathcal B(n)$ is relex ordered. This removes the necessity to check whether {\bf a} $\succcurlyeq \bhash$ from then on.
\newpage
\begin{algorithm}
\SetDataSty{smallsf}
\SetFuncSty{smallsf}
\SetKwSty{bfsf}
\SetKw{Downto}{downto}
\SetKw{From}{from}
\SetKw{In}{in}
\SetKwProg{Function}{Function}{}{}
\SetKwFunction{RootedTrees}{RootedTrees}
\SetKwFunction{RTHelper}{RTHelper2}
\SetKwData{Bqn}{Bqn}
\SetKw{Break}{break}
\SetKwData{a}{a}
\SetKwData{b}{b}
\SetKwData{RTHelperList}{RTHelperList}
\SetAlgoNoEnd\SetAlgoNoLine\DontPrintSemicolon
\setstretch{1.1}
\Function{\RTHelper{$n, q$}}
{
\lIf{$q=1$}{\Return $n \, \smalloplus\, 1^{n-1}$}
\Bqn \la [ ]\;
\uIf{$q=2$}
	{\For{$t$ \From $\left\lfloor \frac{1}{2}(n-1) \right\rfloor$ \Downto $1$}
		{\Bqn \la \Bqn $\oplus$ [$n \, \smalloplus \, (2\,  1)^t \smalloplus 1^{n - 1 - 2t}]$}
	}
\uElseIf{$q=n-1$}
	{\lFor{\a \In \RootedTrees{$q$}}{\Bqn \la \Bqn $\oplus \, [n \, \smalloplus$ \a]}
	}
\uElseIf{$q=n-2$}
	{\lFor{\a \In \RootedTrees{$q$}}
	{\Bqn \la \Bqn $\oplus \, [n \, \smalloplus$ \a $\smalloplus \, 1$]}}
\Else
	{$newq$ \la $\min(n - q - 1, q)$\;
	\For{\a \In \RootedTrees{$q$}}
		{\For{$r$ \From $newq$ \Downto $1$}
			{\RTHelperList \la \RTHelper{$n - q, r$}\;
			$start$ \la $1$\;
			\If{$r=q$}
				{\For{\b \In \RTHelperList}
					{\lIf{\a $\succcurlyeq$  \b$^{\!\!\#}$}{\Break}
					{$start$ \la $start + 1$\;}}
				}
			\lFor{\b \In \RTHelperList$\!\![start \ldots ]$}{\Bqn \la  \Bqn $\oplus$ [$n\, \smalloplus$ {\a} $\smalloplus$ \b$^{\!\!\#}$]}
			}
		}
	}
{\Return \Bqn}
}
\end{algorithm}
We shall assume from now on that the functions \smallsf{RootedTrees} and \smallsf{UFT} use the helper function \smallsf{RTHelper2} instead of \smallsf{RTHelper}.\vspace{-2.5mm}
\subsection{\normalsize Caching of $\mathcal B(k)$ for smaller values of $k$}\label{caching B(k)}\vspace{-2.5mm}
We now discuss how we can further improve the efficiency of the tree generation algorithms by caching $\mathcal B(k)$ for small values of $k$. 

\smallsf{RTHelper2}$(n,\, q)$ calls \smallsf{RootedTrees}$(q)$ and \smallsf{RTHelper2}$(n - q,\, r)$, where $r \le q$, and \smallsf{RootedTrees}$(q)$ calls \smallsf{RTHelper2}$(q,\, q^\prime)$, where $q^\prime < q$. It follows that $n^\prime \le q$ in all the calls to \smallsf{RootedTrees}$(n^\prime)$ made by \smallsf{RTHelper2}$(n,\, q)$, whether directly or indirectly. It therefore follows that we can obtain a significant increase in the efficiency of the function call \smallsf{RTHelper2}$(n, q)$ if we cache in memory $\mathcal B(k)$, for $1 \le k \le q$. This will increase the efficiency of both \smallsf{RootedTrees} and \smallsf{UFT}.

For rooted tree generation, \smallsf{RootedTrees}$(n)$ makes calls to \smallsf{RTHelper2}$(n,\, q)$, where $1 \le q \le n - 1$. However, for large values of $n$, the space requirements to cache $\mathcal B(k)$, for $1 \le k \le n-1$, would be prohibitive. For unicentroidal free tree generation, $q \le \left\lfloor\frac{n-1}{2}\right\rfloor$ for all calls to \smallsf{RTHelper2}$(n,\, q)$ made by \smallsf{UFT}$(n)$. Furthermore, since \smallsf{BFT}$(n)$ only makes calls to \smallsf{RootedTrees}$(\frac{n}{2})$, caching $\mathcal B(k)$ for $1 \le k \le \frac{n}{2}$ would avoid all calls of \smallsf{RootedTrees} for both \smallsf{UFT}$(n)$ and \smallsf{BFT}$(n)$, and thus also \smallsf{FreeTrees}$(n)$. So, for example, to generate all $109,972,410,221$ free trees of order $32$, this would mean caching $\mathcal B(k)$ for $1 \le k \le 16$, which is perfectly feasible since there only $235,381$ rooted trees of order $16$. On the other hand, in order to generate all {\em rooted} trees of order $32$ whilst avoiding all calls of \smallsf{RootedTrees}, we would need to cache $\mathcal B(k)$ for $1 \le k \le 31$. Since there are nearly $10^{12}$ rooted trees of order $31$, the cache space requirements for generating all rooted trees of order $32$ would be of the order of at least $10$ terabytes, and thus infeasible on practically all current computers (see \cite{oeis} and \cite{rootedoeis} for tree counts).

\newpage
Suppose that we cache $\mathcal B(k)$ for $1 \le k \le L$, and store the sets $\mathcal B(1), \, \mathcal B(2), \ldots, \mathcal B(L)$ in the list \smallsf{RTList}. We can then replace all calls of \smallsf{RootedTrees}$(q)$ in \smallsf{RTHelper2}$(n, q)$ by references to \smallsf{RTList}$[q]$, provided $L \ge q$. 

As there are very few rooted trees of order less than five, we explicitly create $\mathcal B(1)$, $\mathcal B(2)$, $\mathcal B(3)$ and $\mathcal B(4)$ before calling \smallsf{RootedTrees}($k$) for larger values of $k$, as follows:

\begin{algorithm}
\SetDataSty{smallsf}
\SetFuncSty{smallsf}
\SetKwSty{bfsf}
\SetKw{From}{from}
\SetKw{To}{to}
\SetKwProg{Procedure}{Procedure}{}{}
\SetKwFunction{RootedTrees}{RootedTrees}
\SetKwFunction{InitialiseRTList}{InitialiseRTList}
\SetKwData{RTList}{RTList}
\SetAlgoNoEnd\SetAlgoNoLine\DontPrintSemicolon
\setstretch{1.1}
\Procedure{\InitialiseRTList{$L$}}
{
\RTList$\!\![1]=[1]$\;
\RTList$\!\![2]=[2\, 1]$\;
\RTList$\!\![3]=[3\, 2\, 1, \, 3\, 1\, 1]$\;
\RTList$\!\![4]=[4\, 3\, 2\, 1, \, 4\, 3\, 1\, 1, \, 4\, 2\, 1\, 1, \, 4\, 1\, 1\, 1]$\;
\lFor{$k$ \From $5$ \To $L$}
	{\RTList$\!\![k]$ \la \RootedTrees{$k$}}
}
\end{algorithm}

In this initialisation, when computing \smallsf{RootedTrees}$(k)$, we note that we will already have computed the previous elements of \smallsf{RTList}; so the recursive calls of \smallsf{RootedTrees} in \smallsf{RTHelper2} may be replaced by references to \smallsf{RTList}. 

As explained above, we can replace all calls to \smallsf{RootedTrees} from \smallsf{FreeTrees}$(n)$ by references to \smallsf{RTList} if $L \ge \left\lfloor\frac{n}{2}\right\rfloor$. In practice, as explained below, in order to improve the efficiency of the code when $q=\left\lfloor\frac{n-1}{2}\right\rfloor$, we henceforth assume that $L \ge \left\lfloor\frac{n}{2}\right\rfloor+1$.

To avoid all calls to \smallsf{RootedTrees}$(q)$ in \smallsf{RTHelper2}$(n, q)$, we require that $L \ge q$. This is clearly always true for \smallsf{FreeTrees}. We will see that we can also avoid the recursive calls to \smallsf{RTHelper2}$(n - q, r)$ when $q \ge n - L$.

When $q$, the order of the first subtree of the root, is at least $\left\lfloor\frac{n-1}{2}\right\rfloor$, then $newq=n-q-1 \le L$. So the $r$-loop (where $r$ is the order of the second subtree of the root) can be dispensed with by letting {\bf b} iterate through \smallsf{RTList}$[n - q]$. Now, when $q$ is at least $\left\lfloor\frac{n+1}{2}\right\rfloor$, then $r \le newq < q$, so we can dispense with checking whether ${\bf a} \succcurlyeq {\bf b^{\#}}$. When $q = \left\lfloor\frac{n-1}{2}\right\rfloor$ and $r=q$, we can also avoid checking whether ${\bf a} \succcurlyeq {\bf b^{\#}}$ by skipping the initial elements of \smallsf{RTList$[n - q]$}, as we now explain.

Suppose that $r =q = \left\lfloor\frac{n-1}{2}\right\rfloor$ and ${\bf a} =  \smallsf{RTList}[q][k]$. When $n$ is odd, $n-q-1=q$, so we can start with the element {\bf b} for which ${\bf b^{\#}}= {\bf a}$; this is easily seen to be $\smallsf{RTList}[n - q][k]$. When $n$ is even, $n-q-1=q+1=\f{n}{2}$, so we can skip the $|\mathcal B(\frac{n}{2})|$ elements for which the first subtree is of order $\frac{n}{2}$, and start with the element {\bf b} for which ${\bf b^{\#}} = {\bf a} \smalloplus 1$; this is easily seen to be $\smallsf{RTList}[n - q][k + s]$, where $s=|\mathcal B(\frac{n}{2})|$.

When $\left\lfloor\frac{n-1}{2}\right\rfloor > q \ge n - L$, we can again dispense with the $r$-loop, letting {\bf b} iterate through part of the \smallsf{RTList}$[n - q]$. We pre-compute an array \smallsf{RTqstart}, where \smallsf{RTqstart}$[n][q]$ is the index of the first element in the list \smallsf{RTList}$[n]$ for which the first subtree of the root of the corresponding rooted tree is of order $q$. We store in memory \smallsf{RTqstart}$[n][q]$, where $1 \le q < n \le L$. Since ${\bf a} \succcurlyeq \bhash$ cannot hold for any element ${\bf b}$ in \smallsf{RTList}$[n - q]$ with index less than \smallsf{RTqstart}$[n - q][q]$, we only need to consider those elements ${\bf b}$ in \smallsf{RTList}$[n - q]$ from this index on.

We note that, following the above changes, we can replace $newq$ by $q$ when $q < n - L$, since $q \le L$. Making these changes to \smallsf{RTHelper2} yields the function \smallsf{RTHelper3}. We shall assume from now on that the functions \smallsf{RootedTrees} and \smallsf{UFT} use the helper function \smallsf{RTHelper3} instead of \smallsf{RTHelper2}.

\begin{algorithm}
\SetDataSty{smallsf}
\SetFuncSty{smallsf}
\SetKwComment{NC}{   \#   }{}
\SetKwSty{bfsf}
\SetKw{Downto}{downto}
\SetKw{From}{from}
\SetKw{In}{in}
\SetKwProg{Function}{Function}{}{}
\SetKwFunction{RTHelper}{RTHelper3}
\SetKwFunction{Length}{length}
\SetKwData{Bqn}{Bqn}
\SetKwData{RTQstart}{RTqstart}
\SetKwData{a}{a}
\SetKwData{b}{b}
\SetKwData{RTHelperList}{RTHelperList}
\SetKwData{RTList}{RTList}
\SetKwData{Break}{break}
\SetKwData{PrunedList}{PrunedList}
\SetAlgoNoEnd\SetAlgoNoLine\DontPrintSemicolon
\setstretch{1.1}
\Function(\NC*[h]{This assumes that $L \ge q$ and $L \ge \left\lfloor\frac{n}{2}\right\rfloor+1$.}){\RTHelper{$n, q$}}
{
\lIf{$q=1$}{\Return $n \, \smalloplus\, 1^{n-1}$}
\Bqn \la [ ]\;
\uIf{$q=2$}
	{\For{$t$ \From $\left\lfloor \frac{1}{2}(n-1) \right\rfloor$ \Downto $1$}
		{\Bqn \la \Bqn $\oplus [n \, \smalloplus \, (2\,  1)^t \smalloplus 1^{n - 1 - 2t}]$\;}
	}
\uElseIf{$q=n-1$}
	{\lFor{\a \In \RTList$\!\![q]$}{\Bqn \la \Bqn $\oplus \, [n \, \smalloplus$ \a]}}
\uElseIf{$q=n-2$}
	{\lFor{\a \In \RTList$\!\![q]$}{\Bqn \la \Bqn $\oplus \, [n \, \smalloplus$ \a $\smalloplus \, 1$]}}
\uElseIf{$q \ge  \left\lfloor\frac{n+1}{2}\right\rfloor$}
	{\For{\a \In \RTList$\!\![q]$}
		{\lFor{\b \In \RTList$\!\![n-q]$}{\Bqn \la  \Bqn $\oplus$ [$n\, \smalloplus$ {\a} $\smalloplus$ \b$^{\!\!\#}$]}}
	}
\uElseIf{$q =  \left\lfloor\frac{n-1}{2}\right\rfloor$}
	{$start$ \la $1$\;
	\lIf{n $\equiv 0 \pmod 2$}{$start$ \la \Length{\RTList$\!\![\frac{n}{2}]$} + 1}
	\For{\a \In \RTList$\!\![q]$}
		{\lFor{\b \In \RTList$\!\![n-q][start \ldots]$}{\Bqn \la  \Bqn $\oplus$ [$n\, \smalloplus$ {\a} $\smalloplus$ \b$^{\!\!\#}$]}
		$start$ \la $start + 1$}
	}
\uElseIf{$q \ge n - L$}
	{$start$ \la \RTQstart$\!\![n-q, q]$\;
	\For{\a \In \RTList$\!\![q]$}
		{\For{\b \In \RTList\!\!$[n-q][start \ldots ]$}
			{\lIf{\a $\succcurlyeq$ \b$^{\!\!\#}$}{\Break}
			$start$ \la $start + 1$\;}
		\lFor{\b \In \RTList$\!\![n-q][start \ldots ]$}{\Bqn \la  \Bqn $\oplus$ [$n\, \smalloplus$ {\a} $\smalloplus$ \b$^{\!\!\#}$]}
		}
	}
\Else
	{\For{\a \In \RTList$\!\![q]$}
		{\For{$r$ \From $q$ \Downto $1$}
			{\RTHelperList \la \RTHelper{$n - q, r$}\;
			$start$ \la $1$\;
			\If{$r=q$}
				{\For{\b \In \RTHelperList}
					{\lIf{\a $\succcurlyeq$ \b$^{\!\!\#}$}{\Break}
					$start$ \la $start + 1$\;
					}
				}
			\lFor{\b \In \RTHelperList$\!\![start \ldots ]$}{\Bqn \la  \Bqn $\oplus$ [$n\, \smalloplus$ {\a} $\smalloplus$ \b$^{\!\!\#}$]}
			}
		}
	}
{\Return \Bqn}
}
\end{algorithm}
\newpage
In practice, as well as caching the $\mathcal B(k)$ in \smallsf{RTList}(k), we also cache, in \smallsf{RFList}(k), the relex ordered set of sequences that is obtained by replacing each sequence {\bf b} in $\mathcal B(k)$ by ${\bf b^{\#}}$. We can then make ${\bf b^{\#}}$ iterate through the corresponding forest sequences instead of {\bf b} iterating through the tree sequences, obviating the need to remove the first element of {\bf b} each time. We therefore return ${\bf a} \smalloplus {\bf b^{\#}}$ instead of $n \smalloplus {\bf a} \smalloplus {\bf b^{\#}}$ in the function \smallsf{RFHelper} that generates the sequence ${\bf s^{\#}}$ for each sequence ${\bf s}$ generated by \smallsf{RTHelper3}. We prepend the weight of the root to each forest in the functions \smallsf{RootedTrees} and \smallsf{UFT}. Although we have not included the pseudocode for \smallsf{RFHelper}, we use this function in our Python implementations instead of \smallsf{RTHelper3}.

We note that, we may improve the efficiency of the function \smallsf{BFT} by using the same idea as that used for the case $q= \left\lfloor\frac{n-1}{2}\right\rfloor$ in the function \smallsf{RTHelper3}. This yields the following function \smallsf{BFT2}.

\begin{algorithm}
\SetDataSty{smallsf}
\SetFuncSty{smallsf}
\SetKwSty{bfsf}
\SetKw{Downto}{downto}
\SetKw{From}{from}
\SetKw{In}{in}
\SetKwProg{Function}{Function}{}{}
\SetKwFunction{BFT}{BFT2}
\SetKwData{RTList}{RTList}
\SetKwData{BFn}{BFn}
\SetKwData{a}{a1}
\SetKwData{aa}{a2}
\SetAlgoNoEnd\SetAlgoNoLine\DontPrintSemicolon
\setstretch{1.1}
\Function{\BFT{$n$}}
{
\BFn \la [ ]\;
{$start$ \la $1$\;
\For{\a \In \RTList\!\!$[\frac{n}{2}]$}
	{\lFor{\aa \In \RTList\!\!$[\frac{n}{2}][start ...]$}{\BFn \la \BFn $\oplus$ [\a $\smalloplus$ \aa]}
	$start$ \la $start + 1$\;}
\Return \BFn}
}
\end{algorithm}
We shall assume from now on that the function \smallsf{FreeTrees} uses the function \smallsf{BFT2} instead of \smallsf{BFT}.\vspace{-2.5mm}
\subsection{\normalsize Generators}\label{generators}\vspace{-2.5mm}
The size of $\mathcal B(n)$ grows exponentially, so the list \smallsf{Bqn} may become prohibitively large for large values of $n$, except when $q$ is small. Therefore, to avoid creating and returning the list \smallsf{Bqn} in \smallsf{RTHelper3}, we instead return a generator. The changes necessary to effect this are, in essence, to simply replace all the assignments of the form \smallsf{Bqn} \la \smallsf{Bqn} $\oplus$ [{\bf c}] by the statement \smallsf{yield} {\bf c}, and make corresponding changes to the other algorithms.\vspace{-2.5mm}
\subsection{\normalsize Strings for sequences}\label{strings for sequences}\vspace{-2.5mm}
We store the weight sequences of the trees as alphanumeric strings, instead of lists, both to save storage and to create the canonical weight sequences more efficiently. We use the digits $1$ to $9$ for the corresponding weights, and the letters $A, \, B, \, C, \ldots $ for weights $10, \,11,\, 12, \ldots$. So the weight sequence of the free tree $T$ in Figure \ref{fig 3} is denoted by the string $``A421141111"$ instead of the sequence $10\,\, 4\,2\,1\,1\, 4\,1\,1\,1\,1$.
\subsection{\normalsize Adjacency lists and matrices}\label{adjacency lists and matrices}
Although weight sequences are useful for generating trees, for most purposes a more conventional representation is required, such as adjacency lists or adjacency matrices. Most other tree generation algorithms also initially generate the trees using non-conventional  representations (e.g., level sequences or parent sequences, as mentioned in the introduction). The adjacency lists or matrices are then constructed from the particular representation used.

We now give a brief explanation of how we can incorporate the construction of the adjacency lists of the free trees of order $n$ into our algorithm \smallsf{FreeTrees}, using a caching approach similar to that outlined in Section \ref{caching B(k)}. We assume that the vertices are labelled $1$ to $n$ in preorder.

The algorithm \smallsf{AdjListFromWS} below returns the adjacency list of a single free tree given its weight sequence. In the algorithm, we denote the $j^{th}$ element of the weight sequence \smallsf{ws} by \smallsf{ws}$[j]$, and the list of $n$ empty lists by $[\, ]^n$.

\begin{algorithm}
\SetDataSty{smallsf}
\SetFuncSty{smallsf}
\SetKwSty{bfsf}
\SetKw{To}{to}
\SetKw{From}{from}
\SetKw{In}{in}
\SetKwProg{Function}{Function}{}{}
\SetKwFunction{AdjListWtSeq}{AdjListFromWS}
\SetKwFunction{Length}{length}
\SetKwData{A}{A}
\SetKwData{ws}{ws}
\SetAlgoNoEnd\SetAlgoNoLine\DontPrintSemicolon
\setstretch{1.1}
\Function{\AdjListWtSeq{\ws}}
{
$n$ \la \Length{\ws}\;
\A \la [ ]$^n$\;
\For{$i$ \From $1$ \To $n$}
{$j=i+1$\;
\While{$j < i \, + $ \ws $\!\![i]$}
{\A$\!\![i]$ \la \A$\!\![i] \oplus [j]$\;
\A$\!\![j]$ \la \A$\!\![j] \oplus [i]$\;
$j$ \la $j\, + \, $\ws $\!\![j]$}}
\If{\ws $\!\![1]= \f{n}{2}$}
{$hn=\f{n}{2}\, + \, 1$\;
\A$\!\![1]$ \la \A$\!\![1] \oplus [hn]$\;
\A$\!\![hn]$ \la \A$\!\![hn] \oplus [1]$\;}
\Return \A
}
\end{algorithm}

We note that, given the weight sequence of any {\em ordered} tree (or indeed any ordered forest), whether canonically ordered or not, this algorithm will return its adjacency list if we remove the assignment \smallsf{A}$[j]$ \la \smallsf{A}$[j]  \oplus [i]$ and the \smallsf{if} statement (which, for a bicentroidal tree, adds the edge between the two centroids).

We extend the procedure \smallsf{InitialiseRTList} to construct the adjacency list representations of the rooted trees, by calling the function \smallsf{AdjListFromWS} on each weight sequence in \smallsf{RTList}[$k$]. We store these representations in a hash table (implemented as a Python dictionary) using the weight sequence as the key.

We can now construct the adjacency list representation of all the free trees of order $n$ while we construct their weight sequences: for each subtree of the root, we look up its adjacency list representation in the hash table, and then increase the label of each vertex by a suitable offset value. For a unicentroidal free tree represented by the integer sequence $n \smalloplus \, {\bf a} \, \smalloplus \, {\bf b^{\#}}$, we offset the labels of the vertices of the subtree correponding to {\bf a} by $1$, and those of the vertices of the forest corresponding to ${\bf b^{\#}}$ by $|{\bf a}| +1$. For a bicentroidal free tree, we only need to offset the labels of the vertices corresponding to the subtree rooted at the second bicentroid by $\f{n}{2}$. 

It is fairly straightforward to modify the above procedure in order to generate adjacency matrices instead of adjacency lists in a similar manner. The Python code for generating both the adjacency list and matrix representations is included in the appendix.
\vspace{-2.5mm}
\section{\normalsize Free tree generation runtimes and comparisons}\label{time tests}\vspace{-2.5mm}

As $n$, the order of the trees we are generating, is never large, we assume a uniform cost criterion, i.e., that the string operations, such as concatentation, take constant time. It is straightforward to prove, by induction, that the runtimes of \smallsf{RootedTrees}$(n)$ and \smallsf{RTHelper}$(n, q)$ would be at most proportional to the number of sequences returned by these functions if we removed from \smallsf{RTHelper} the test ${\bf a} \succcurlyeq {\bf b}^{\#}$. We have computed empirically that the additional time taken for scanning along the list until ${\bf a} \succcurlyeq {\bf b}^{\#}$ is equivalent to increasing the runtime by less than 5\%. Therefore, under the uniform cost criterion assumption, the generation time per tree would be approximately constant.
\newpage
We now present an empirical comparison of our algorithm with the popular WROM algorithm. We implemented our algorithms in Python and compared these with the Python implementation of the WROM algorithm taken from NetworkX. All computations were performed using Python $3.7$ and the JIT compiler PyPy$3.6$-v$7.3.1$, running on a Pentium i7 with 16GB RAM. We set $L$, the order of the largest tree for which we cache the representations, to be $\f{n}{2}+1$.

Table \ref{run times} shows the times in seconds to generate all free trees of order $n$ and return the count of the number of trees, without saving the representations. BRFE refers to the algorithm \smallsf{FreeTrees} described above and WROM to the algorithm described in \cite{wrom}. BRFE(ls) and BRFE(mat) include converting the weight sequences into the adjacency list and matrix representations, respectively; WROM(ls) and WROM(mat) are defined similarly.

\begin{table}[h]
\begin{tabular}{ |c|c|c|c|c|c|c|c|c| } 
 \hline
$n$&No. of trees&BRFE&BRFE(ls)&BRFE(mat)&WROM&WROM(ls)&WROM(mat)\\
 \hline \hline
$18$&$123867$&$0.05$&$0.21$&$0.18$&$0.22$&$0.33$&$0.38$\\
$19$&$317955$&$0.08$&$0.29$&$0.26$&$0.29$&$0.53$&$0.69$\\
$20$&$823065$&$0.12$&$0.47$&$0.44$&$0.53$&$1.14$&$1.58$\\
$21$&$2144505$&$0.28$&$0.93$&$0.84$&$1.10$&$2.62$&$3.82$\\
$22$&$5623756$&$0.53$&$2.02$&$1.86$&$2.64$&$6.79$&$10.12$\\
$23$&$14828074$&$1.61$&$5.33$&$4.66$&$6.50$&$17.95$&$27.34$\\
$24$&$39299897$&$3.41$&$13.56$&$12.34$&$17.72$&$48.41$&$75.49$\\
$25$&$104636890$&$11.09$&$40.47$&$37.29$&$49.73$&$129.57$&$207.03$\\
$26$&$279793450$&$24.24$&$104.66$&$102.16$&$127.31$&$364.85$&$594.39$\\
$27$&$751065460$&$80.26$&$319.91$&$288.38$&$336.57$&$-$&$-$\\
$28$&$2023443032$&$174.68$&$-$&$-$&$-$&$-$&$-$\\
$29$&$5469566585$&$657.46$&$-$&$-$&$-$&$-$&$-$\\
 \hline
\end{tabular}
\caption{Runtimes in seconds of the implementations of the BRFE and WROM algorithms}\label{run times}
\end{table}

As can be seen, the runtimes for generating the weight sequences using BRFE are less than a quarter of those for generating the level sequences using WROM. The speed-ups for the times to create the adjacency list and matrix representations are similar. Due to the excessive times involved, we have not run some of the algorithms for the larger values of $n$.

We note that the runtimes for BRFE are about four times as long using the standard CPython implementation as those in Table \ref{run times}, and the run times for WROM are about ten times as long. We further note that, by increasing the value of $L$, we could significantly reduce the runtimes of our algorithms for larger values of $n$, e.g., the runtime when $n=29$ and $L=19$ is around $556$ seconds.

Li and Ruskey presented an alternative algorithm in \cite{li} \cite{lirusk} that generates parent sequences, and compared a PASCAL implementation of their algorithm and the WROM algorithm. It can be seen from Table $5.2$ in \cite{li} that the runtime of their algorithm is about $70\%$ of that of WROM. We can deduce from this that BRFE would take about a third of the time of their algorithm.

In Table \ref{run times per tree}, we show the corresponding generation times per tree in nanoseconds. For the smaller values of $n$, the times are inflated by the start-up times of the PyPy$3$ JIT compiler. These times seem to be approximately constant for $n \ge 22$, although for the BFRE algorithms there is some additional fluctuations between odd and even $n$, as the bicentroidal tree algorithm is faster than the unicentroidal. The increases in the times for the largest values of $n$ for the BRFE algorithms are probably due to an increase in the number of hardware cache misses.

\newpage
Although appoximately constant, there is an increase in the times in Table \ref{run times per tree} as $n$ increases, presumably because of the over-simplification in assuming a uniform cost criterion. If we assume the cost of string/list/matrix operations is proportional to the length, and divide the times in Table \ref{run times per tree} by $n$, we obtain the times in Table \ref{run times ratio}. As we might expect, the times are now generally decreasing with $n$, since many of the operations involve smaller strings/lists/matrices. 

\begin{table}
\begin{tabular}{ |c|c|c|c|c|c|c|c|c| } 
 \hline
$n$&No. of trees&BRFE&BRFE(ls)&BRFE(mat)&WROM&WROM(ls)&WROM(mat)\\
 \hline \hline
$18$&$123867$&$403.66$&$1695.37$&$1453.17$&$1776.10$&$2664.15$&$3067.81$\\
$19$&$317955$&$251.61$&$912.08$&$817.73$&$912.08$&$1666.90$&$2170.12$\\
$20$&$823065$&$145.80$&$571.04$&$534.59$&$643.93$&$1385.07$&$1919.65$\\
$21$&$2144505$&$130.57$&$433.67$&$391.70$&$512.94$&$1221.73$&$1781.30$\\
$22$&$5623756$&$94.24$&$359.19$&$330.74$&$469.44$&$1207.38$&$1799.51$\\
$23$&$14828074$&$108.58$&$359.45$&$314.27$&$438.36$&$1210.54$&$1843.80$\\
$24$&$39299897$&$86.77$&$345.04$&$314.00$&$450.89$&$1231.81$&$1920.87$\\
$25$&$104636890$&$105.99$&$386.77$&$356.38$&$475.26$&$1238.28$&$1978.56$\\
$26$&$279793450$&$86.64$&$374.06$&$365.13$&$455.01$&$1304.00$&$2124.39$\\
$27$&$751065460$&$106.86$&$425.94$&$383.96$&$448.12$&$-$&$-$\\
$28$&$2023443032$&$86.33$&$-$&$-$&$-$&$-$&$-$\\
$29$&$5469566585$&$120.20$&$-$&$-$&$-$&$-$&$-$\\
 \hline
\end{tabular}
\caption{Generation times per tree in nanoseconds of the implementations of the algorithms}\label{run times per tree}
\end{table}
\begin{table}
\begin{tabular}{ |c|c|c|c|c|c|c|c|c| } 
 \hline
$n$&No. of trees&BRFE&BRFE(ls)&BRFE(mat)&WROM&WROM(ls)&WROM(mat)\\
 \hline \hline
$18$&$123867$&$22.43$&$94.19$&$80.73$&$98.67$&$148.01$&$170.43$\\
$19$&$317955$&$13.24$&$48.00$&$43.04$&$48.00$&$87.73$&$114.22$\\
$20$&$823065$&$7.29$&$28.55$&$26.73$&$32.20$&$69.25$&$95.98$\\
$21$&$2144505$&$6.22$&$20.65$&$18.65$&$24.43$&$58.18$&$84.82$\\
$22$&$5623756$&$4.28$&$16.33$&$15.03$&$21.34$&$54.88$&$81.80$\\
$23$&$14828074$&$4.72$&$15.63$&$13.66$&$19.06$&$52.63$&$80.17$\\
$24$&$39299897$&$3.62$&$14.38$&$13.08$&$18.79$&$51.33$&$80.04$\\
$25$&$104636890$&$4.24$&$15.47$&$14.26$&$19.01$&$49.53$&$79.14$\\
$26$&$279793450$&$3.33$&$14.39$&$14.04$&$17.50$&$50.15$&$81.71$\\
$27$&$751065460$&$3.96$&$15.78$&$14.22$&$16.60$&$-$&$-$\\
$28$&$2023443032$&$3.08$&$-$&$-$&$-$&$-$&$-$\\
$29$&$5469566585$&$4.14$&$-$&$-$&$-$&$-$&$-$\\
 \hline
\end{tabular}
\caption{Ratio of the generation times per tree to $n$ in nanoseconds}\label{run times ratio}
\end{table}
\vspace{-2.5mm}
\section{\normalsize Conclusion}\label{conclusion}\vspace{-2.5mm}
In this paper we have presented new canonical representations for ordered, rooted and free trees. We constructed recursive algorithms for generating all rooted trees and all free trees of order $n$ using these representations; each of these algorithms returns a list of the trees generated. We made a number of improvements to the algorithms and their Python implementations, including using generators to avoid having to explicitly construct and store the long lists of trees returned by the recursive calls. Moreover, in order to eliminate many of the recursive calls for small values of $n$, we cached the lists of rooted trees of small order. Our main interest is in the generation of free trees and, in this case, in order to eliminate a large proportion of the recursive calls, it is only necessary to cache the lists of rooted trees up to order around $\f{n}{2}$. We then described how the algorithm could be modified to generate the adjacency list or matrix representations of the trees.

We compared our Python implementation of the algorithm for generating free trees with the Python implementation of the well-known WROM algorithm taken from NetworkX. We used our algorithm to generate the free trees of order $n$, for $18 \le n \le 29$, but because of the longer runtimes, we only ran the WROM algorithm up to $n=27$. It can be seen from Table \ref{run times} that the runtimes for the new algorithm are less than a quarter of those for the WROM algorithm (the improvement in the runtimes for the algorithms that generate adjacency lists or matrices is similar ). From the comparisons in \cite{li}, we may deduce that our algorithm would take less than a third of the time of the algorithm presented there.

\end{document}